\documentclass[prb,twocolumn,groupedaddress]{revtex4}
\usepackage{amsmath}
\usepackage{amssymb}
\usepackage{amsfonts}
\usepackage[dvips]{graphicx}
\usepackage[]{caption}
\usepackage[]{epsfig}
\newcommand {\Tp}{{\rm T}_+}
\newcommand {\Tm}{{\rm T}_-}
\newcommand {\Tn}{{\rm T}_0}

\newcommand {\Cn}{{\rm C}_0}
\newcommand {\Cp}{{\rm C}_+}
\newcommand {\Cm}{{\rm C}_-}

\newcommand {\Sn}{{\rm S}_0}
\newcommand {\Sp}{{\rm S}_+}
\newcommand {\Sm}{{\rm S}_-}

\begin{document}

\title{Competition of hydrophobic and Coulombic interactions between nano-sized solutes}

\author{J. Dzubiella}
\email[e-mail address:] {jd319@cam.ac.uk}
\affiliation{University Chemical Laboratory,
Lensfield Road,
Cambridge CB2 1EW,
United Kingdom}
\author{J.-P. Hansen}
\affiliation{University Chemical Laboratory,
Lensfield Road,
Cambridge CB2 1EW,
United Kingdom}
\date{\today}

\begin{abstract}
The solvation of charged, nanometer-sized spherical solutes in water,
and the effective, solvent-induced force between two such solutes are
investigated by constant temperature and pressure Molecular Dynamics
simulations of model solutes carrying various charge patterns. The
results for neutral solutes agree well with earlier findings, and with
predictions of simple macroscopic considerations: substantial
hydrophobic attraction may be traced back to strong depletion
(``drying'') of the solvent between the solutes. This hydrophobic
attraction is strongly reduced when the solutes are uniformly charged,
and the total force becomes repulsive at sufficiently high charge;
there is a significant asymmetry between anionic and cationic solute
pairs, the latter experiencing a lesser hydrophobic attraction. The
situation becomes more complex when the solutes carry discrete (rather
than uniform) charge patterns. Due to antagonistic effects of the
resulting hydrophilic and hydrophobic ``patches'' on the solvent
molecules, water is once more significantly depleted around the
solutes, and the effective interaction reverts to being mainly
attractive, despite the direct electrostatic repulsion between
solutes. Examination of a highly coarse-grained configurational
probability density shows that the relative orientation of the two
solutes is very different in explicit solvent, compared to the
prediction of the crude implicit solvent representation.  The present
study strongly suggests that a realistic modeling of the charge
distribution on the surface of globular proteins, as well as the
molecular treatment of water are essential prerequisites for any
reliable study of protein aggregation.
\end{abstract}


\maketitle
\section{Introduction}
It is a well-known fact of physical chemistry that solvophobic solutes
of similar sizes and shapes tend to attract each other in an
incompatible solvent. Classic examples are the effective attraction
between monomers of polymer coils in poor solvent, which leads to
collapse below the $\Theta$-temperature,\cite{rubinstein} or the
attraction between hydrophobic surfaces in
water.\cite{chandler_review} The effective attraction ultimately leads
to phase separation of the solvent and solute as the concentration of
the latter increases. The solvent-averaged effective interaction (or
potential of mean force) is related to the variation of free energy
upon bringing the two solutes from infinite to a finite separation in
the solvent. The change in free energy has an entropic
component, associated with the reorganization of the solvent molecules
around the two solutes, and an energetic contribution
which accounts for the deficit in attractive interactions between
solvent molecules close to the solutes. There is some analogy between
solvophobic attraction and the well-known depletion interaction
between colloidal particles induced by a depletant like non-adsorbing
polymers.\cite{likos:physrep} In the latter case the depletion
attraction can be essentially understood in terms of excluded volume,
and is hence of entropic origin, while hydrophobic interactions have a
large energetic contribution, associated with the
formation or break up of hydrogen bonds.

It has been recognized that the size of the solute plays an important
role in understanding its solvation energy, and effective
solute-solute attraction.\cite{chandler_review} For solutes of a
characteristic size larger than a few molecular diameters (typically
larger than 1nm), a mechanism first envisioned by Stilinger
\cite{stilinger} is that of solvent dewetting (``drying''), i.e. the
solvent molecules tend to move away from the surface of a large
solute, and form a liquid-gas like interface parallel to the solute
interface.\cite{chandler_review, hummer:prl} The overlap of the drying
zones associated with two large solutes as their surfaces come
together may then give rise to an effective attraction,\cite{lum:jpc}
very much like the depletion mechanism between-colloidal
particles. The mechanism holds, a priori, for any solvent, provided
that it is in a thermodynamic state close to liquid-vapor
coexistence.\cite{huang:jpc} The drying mechanism and resulting
attraction between two plate-like solutes was confirmed by Molecular
Dynamics (MD) simulation of Wallquist and Berne.\cite{wallquist:jpc}
Similar simulations were carried out for two large spherical solutes
in a Lennard-Jones solvent, and attractive solvation force profiles
were determined.\cite{kinoshita,shinto,qin}

However most nano-scale biomolecular solutes, like proteins, carry
electric charges, which make them, at least partly, hydrophilic. The
main objective of the present paper is to investigate the influence of
solute-solute and solute-solvent electrostatic interactions on the
effective, solvent-induced potential of mean force between two
solutes. In order to make contact with earlier work on neutral
hard-sphere solutes, we restrict the present investigation to
spherical solutes of identical radii $R\lesssim 1$nm, but carrying
various surface charge patterns. The two main questions which will be
addressed are: a) how does the competition between hydrophobicity and
electrostatics affect the total effective force between anionic and
cationic solutes; b) is the total force sensitive to details of the
charge patterns carried by the solutes? In particular are there
significant differences between results obtained with continuous and
discrete charge patterns of the solutes?  Such differences were
recently highlighted by a calculation of the second virial coefficient
in an implicit solvent model of globular proteins, which does not, of
course, allow for hydrophobic attraction.\cite{elshad:epl:2002} 

We have attempted to answer these questions by a series of constant
pressure MD simulations of two spherical solutes of varying radii (up
to $R=1.3$nm) immersed in water modeled by the SPC/E intermolecular
potential,\cite{berendsen:jpc} taken under normal conditions, i.e. close to
liquid-vapor coexistence. The paper is structured as follows. The
models and simulation procedures are detailed in Sec. \ref{sec:model}. The
solvation of single charged solutes is examined in
Sec. \ref{sec:single}. The effective interaction between two solutes as
a function of the mutual distance is estimated from a simple
macroscopic theory in Sec. \ref{sec:force:theory}, where the method for
extracting the mean effective force from simulations is also
defined. The results from MD simulations for several charge patterns
are presented in Sec. \ref{sec:force:sim}, while concluding remarks are
made in Sec. \ref{sec:conclusion}.

Part of the present results were briefly reported elsewhere.\cite{dzubiella:jcp:2003}

\section{Models and methodology}

\label{sec:model}

\begin{figure}
  \begin{center}
\includegraphics[width=5.0cm,angle=0.,clip]{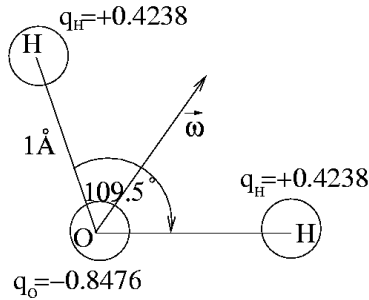}
    \caption{Sketch of the SPC/E water model. The vector $\vec\omega$
    embodies the orientation of the molecule.}
\label{fig:spce} 
\end{center}
\end{figure}

The MD simulations were carried out on periodic samples containing
$N_{\rm w}$ water molecules and one or two solutes.  The SPC/E model
of a water molecule \cite{berendsen:jpc} is sketched in Fig.~1. Two
water molecules interact via a Lennard-Jones potential between the
oxygen (O) sites, and the bare Coulomb potentials between the 9 pairs
of sites. The Lennard-Jones parameters are
$\epsilon=0.6502$kJmol$^{-1}$ and $\sigma=3.169$\AA. The molecules are
assumed to be rigid (with OH bond lengths and HOH bond angle specified
in Fig.~1) and nonpolarizable.  The solutes are smooth spheres of
bare radius $R_{0}$, which interact with the O-site of the water
molecules by the purely repulsive potential
\begin{eqnarray}
V_{0}(r) = \phi (r-R_{0})^{-12},
\label{eq:solutepot}
\end{eqnarray}  where $r$ is the distance from the solute center to
the O-site, and the energy scale $\phi$ is chosen such that the O-atom
experiences a repulsive energy $k_{\rm B}T$ at a distance
$r-R_{0}=$1\AA~ from the solute surface. With this convention, the
effective radius of the solutes may be defined as $R=R_{0}+1$\AA.  The
purely repulsive interaction (\ref{eq:solutepot}) is chosen to mimic a
strongly hydrophobic interaction between the neutral solute and the
water molecules. The model involving spherical solutes with no charged
site will be referred to as $\Sn$.

Since the main objective of our work is to investigate the difference
in the effective, solvent induced interaction between the cases of
neutral and charged solutes, we have considered several models for the
latter (cf. Table 1). In the simplest model, a total charge $Q=qe$
(where $e$ is the proton charge) is assumed to be uniformly
distributed over the solute surface. According to Gauss' theorem, this
is equivalent to placing a single charged site $Q$ at the center of the
spherical solute. We consider both anionic ($q<0$) and cationic
($q>0$) solutes and the corresponding models will be referred to as
$\Sm$ and $\Sp$. To ensure overall electroneutrality of the system,
the total charge carried by the solutes must be compensated either by
a uniform background of total opposite charge permeating the system,
or by explicitly including counterions. For the latter we choose
Cl$^-$ (cationic solutes) and Na$^+$ (anionic solutes) ions. Their
mutual interactions and coupling to water molecules involve the
standard Coulomb-interactions, and a Lennard-Jones part,
with $\epsilon$ and $\sigma$ parameters taken from
Spohr.\cite{spohr:1999} The short range interaction with the solutes
is again described by (\ref{eq:solutepot}).

\begin{table}
\begin{center}
\begin{tabular}{c|c c c|c c c|c c c}
 & $\Sn$ & $\Sp$ & $\Sm$ & $\Tn$ & $\Tp$ & $\Tm$ & $\Cn$ & $\Cp$ & $\Cm$ \\
\hline
$N_c$           & 0   & 1  & 1  & 4 & 4 &  4 & 8  &  8 &  8 \\
$Q/e$           & 0   & q  & -q & 0 & 4 & -4 & 0  &  8 & -8 \\
$R_0/{\rm \AA}$ & $0-12$ & 10 & 10 &10 &10 & 10 & 10 & 10 & 10 \\
$R_c/{\rm \AA}$ & - & 0 & 0 &10 &10 & 10 & 10 & 10 & 10 \\
$R_{cc}/{\rm \AA}$ & - & 0 & 0 &16.33 &16.33 & 16.33 & 11.55 & 11.55 & 11.55 \\
\end{tabular}
\caption{Characteristics of the different models
${\rm S}_{0\pm},{\rm T}_{0\pm},{\rm C}_{0\pm}$ used. $N_c$ is the number of charges
carried by the solute, while $Q$ is the net charge. $R_0$ is
the bare solute radius. By $R_c$ we denote the distance of the
charges to the center of the solute, and $R_{cc}$ quantifies the
nearest neighbor distance between the charges. In the charged
${\rm S}$-models the charge is located in the center of the solutes, while
in the ${\rm T}$ and ${\rm C}$ models the charges are distributed tetrahedrally
and cubically on the sphere surface.  }
\label{tab2}
\end{center}
\end{table}

\begin{figure}
  \begin{center}
\includegraphics[width=7.0cm,angle=0.,clip]{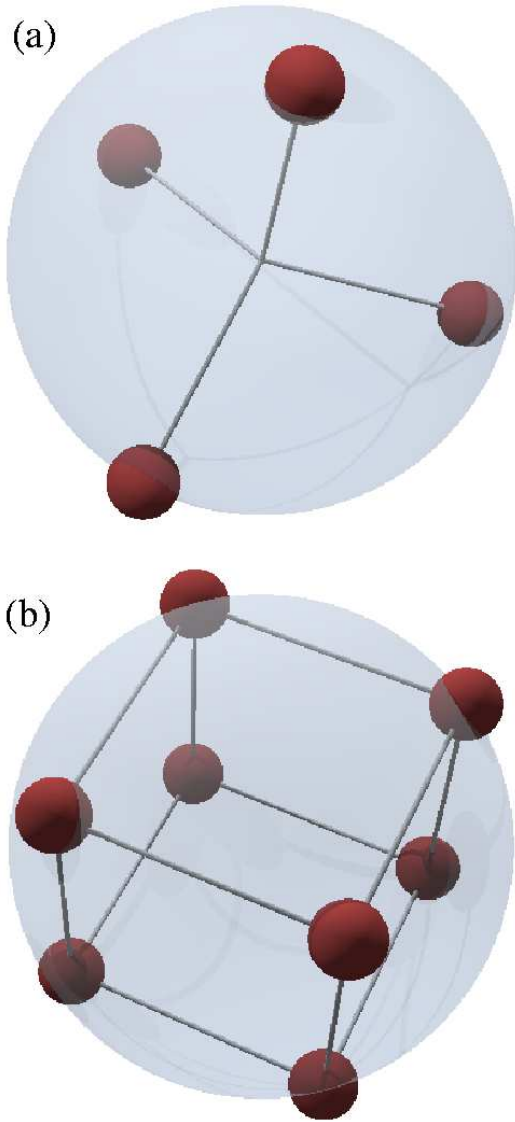}
    \caption{Solute with (a) tetrahedral (T models) and (b) cubic (C
    models) charge distribution. The charges are connected in this
    figure for a better visualization of the structure.}
\label{fig:proteins} 
\end{center}
\end{figure}

In order to investigate the sensitivity of the effective forces to
details of the solute charge patterns, we have also considered models
with discrete charge distributions involving $N_c$ point charges
placed on the surface of the solutes. In the tetrahedron model,
$N_c=4$ charges are tetrahedrally arranged at a distance $R_0$ from
the center of the solutes, as illustrated in
Fig.~\ref{fig:proteins}(a); we consider both the cases where all 4
charges are of the same sign ($q=\pm 4$, referred to as models $\Tp$
and $\Tm$), and the neutral situation, where two charges are positive
and two are negative (model $\Tn$). We have also considered cubic
charge distributions, as sketched in Fig.~\ref{fig:proteins}(b),
where $N_c=8$. In models $\Cp$ and $\Cm$ all 8 charges are positive
($q=8$) or negative ($q=-8$), whereas in model $\Cn$, four vertices
carry a charge $+e$, while the other four carry opposite charges in
an alternating arrangement such that the three nearest neighbors of a
negative charge are positive and vice versa. A similar model of
globular proteins was considered by Allahyarov {\it et
al.},\cite{elshad:epl:2002} but in an implicit (continuous) solvent
representation. 

In order to avoid very close approaches of water
H-sites and the solute surface charges, the charged sites at the
vertices of the tetrahedron or cube, situated at a distance $R_0$ from
the solute center, are not simply point charges, but are modeled by
Cl$^-$ or Na$^+$ ions. The corresponding LJ potentials prevent these
sites and the water H atoms to come too close, and hence unreasonably
large electrostatic forces, which could lead to electrostatic
``sticking'' of the water molecules to the solute surface.
The total interaction energy between one solute and the $N_{\rm w}$ water
molecules in a periodically repeated, cubic simulation cell is:
\begin{eqnarray}
V_{\rm sol} &  = & \sum_{i=1}^{N_{\rm w}} V_{\rm 0}(r_{i})+
\sum_{i=1}^{N_w}\sum_{\alpha=1}^{N_{\rm c}} V_{\rm
  LJ}(r_{i}^{\alpha 1})\nonumber \\
& + &\sum_{i=1}^{N_w}\sum_{\alpha=1}^{N_{\rm c}}\sum_{\beta =1}^3 {q_\alpha}{q_{\beta}}\phi_{\rm EW}(\vec r_{i}^{\alpha\beta}),
\label{eq:totalsolutepot}
\end{eqnarray}
where $r_i$ is the distance from the center of the solute to the
O-atom of the $i$th water molecule, and $r_{i}^{\alpha\beta}$ is the
distance from site $\alpha$ on the solute to site $\beta$ of the $i$th
water molecule ($\beta=1$ for the oxygen site). The first term on the
rhs of Eq.~(\ref{eq:totalsolutepot}) corresponds to the short ranged
repulsion (\ref{eq:solutepot}); the second term is the sum of
Lennard-Jones interactions between all $N_c$ sites of the solute and
the O-sites ($\beta=1$) of the $N_{\rm w}$ water molecules, which
depend on the corresponding site-site distance $r_i^{\alpha 1}$;
finally the last term accounts for the Coulombic interactions between
all $N_c$ solute sites and all 3 water sites; $\phi_{\rm EW}(\vec r)$
is the electrostatic interaction between two elementary charges,
properly summed over an infinite array of periodic images, using the
smooth-particle-mesh Ewald (SPME) method \cite{essmann:jcp} (see the
Appendix A for details). An expression similar to
Eq.~(\ref{eq:totalsolutepot}) holds for the total interaction between a
solute and its Cl$^-$ or Na$^+$ counterions.

The MD simulations were carried out with the DLPOLY2 \cite{dlpoly}
package, using the Verlet leapfrog algorithm,\cite{frenkelsmit} with
a timestep of 2fs. Simulations were carried out at constant pressure
($P=1$atm) and constant temperature ($T=300$K), using appropriate
barostats and thermostats (see Appendix A). We emphasize the importance
of using constant pressure simulations of charged, aqueous systems:
electrostriction and drying mechanisms modify the density of water in
a finite, closed system in a significant way. In the present $NPT$
ensemble simulations, the overall density of water varied from
$\rho_0=0.033{\rm \AA}^{-3}$ to $0.035{\rm \AA}^{-3}$; in estimating
the average water density, the effective volume $4\pi R^3/3$ of the
solutes must be subtracted. The choice of the box length $L$ of the
periodically repeated simulation cell and the arrangement of solutes
inside the cell are discussed in the Appendix A. The cell contained up
to $N_{\rm w}=3000$ water molecules.

\section{Solvation of a single charged solute}
\label{sec:single}

Before embarking on the main subject of this paper, namely the
water-induced effective interaction between two protein-like solutes,
we first consider the solvation of a single, neutral or charged
spherical solute, a problem which has been abundantly addressed in the
literature, since the pioneering work of Born \cite{born} on solvation
free energies of ions, and of Reiss {\it et al.} on the
scaled-particle theory of cavity formation and hard sphere solutes.
\cite{reiss:jcp:1959}

\subsection{Water structure around a solute}
\label{sec:structure}

\begin{figure}
  \begin{center}
\includegraphics[width=8cm,angle=0.,clip]{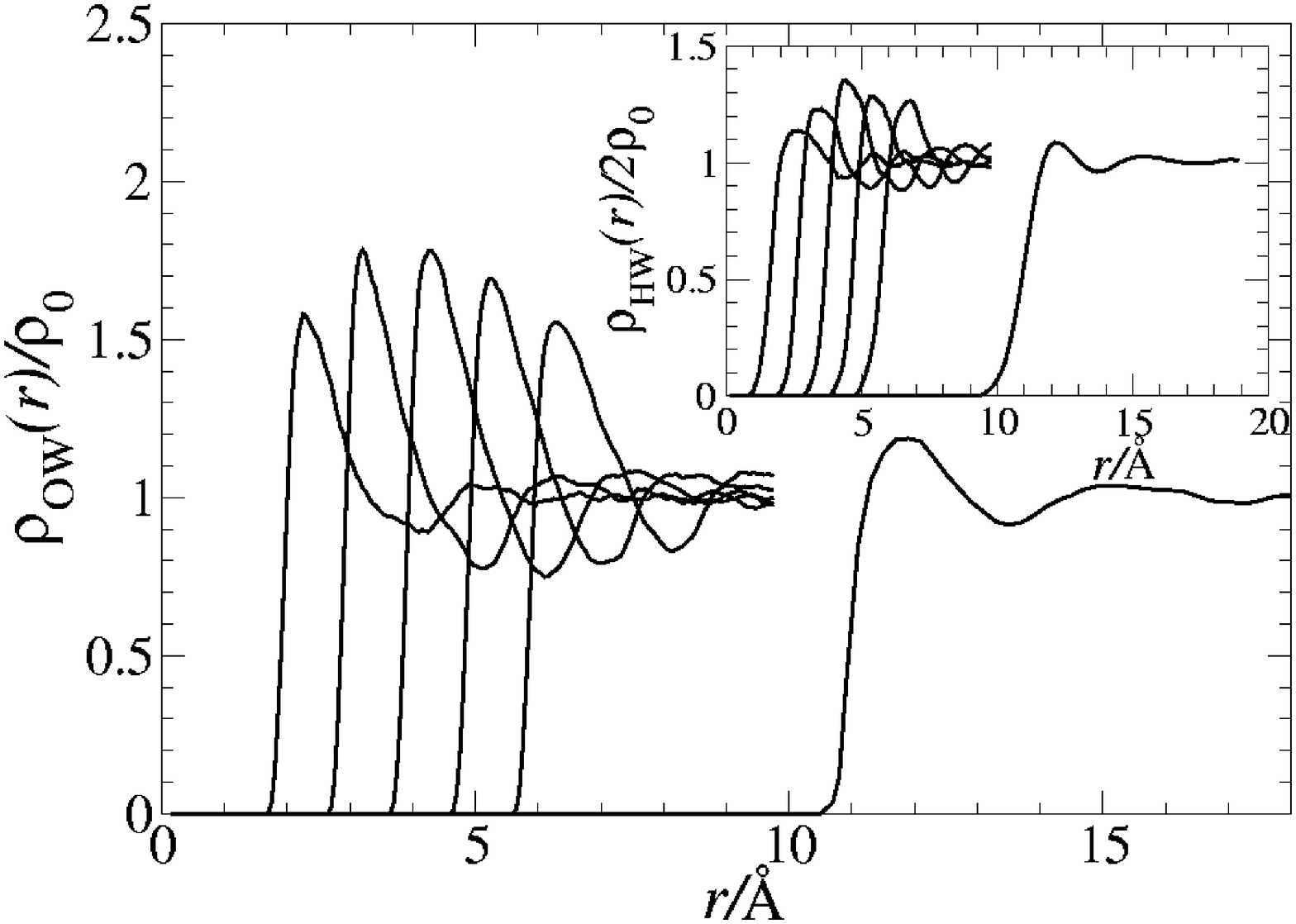}
    \caption{Density profiles of water oxygen and hydrogen atoms
      (inset) around one isolated neutral solute ($\Sn$) plotted for
      different radii $R/{\rm \AA}=2,3,4,5,6,11$.}
\label{fig:profiles_neutral} 
\end{center}
\end{figure}

Consider first the structure of water around an isolated neutral
solute (model $\Sn$). Fig.~\ref{fig:profiles_neutral} shows MD results
for the oxygen and hydrogen radial distribution functions (RDF), or
density profiles, around a solute for various solute radii $R$. The
height of the main peak in the RDF of both O and H sites of the
water molecules is seen to first grow with increasing $R$, and this may
be rationalized in terms of enhanced packing of the molecules at the
surface of the solute. But for $R\gtrsim 5$\AA, the peak height is
seen to decrease monotonically, due to the unbalanced attraction
experienced by the water molecules near the surface from bulk
water. Very similar predictions of the depletion of water around large
spherical solutes (or cavities) have been reported earlier in the
literature.\cite{stilinger,huang:jpc,lum:jpc} The hydrogen and oxygen
peaks are located at nearly the same distance from the solute for any
given radius $R$, with a tendency of the hydrogen peak to be slightly
further out. This seems to indicate that there is no strong
orientation of the water molecules in the first solvation shell
towards or away from the solute surface.

\begin{figure}
  \begin{center}
\includegraphics[width=8cm,angle=0.,clip]{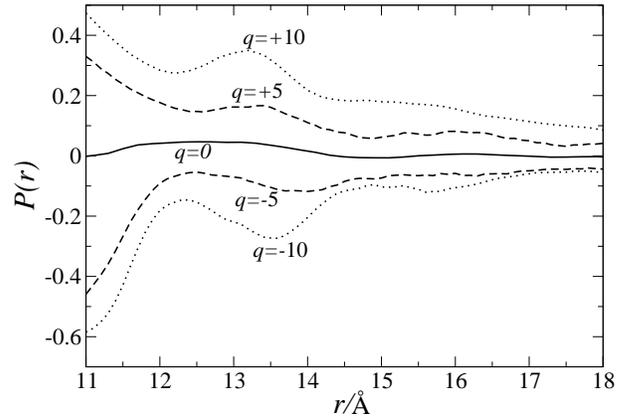}
    \caption{Orientation parameter $P(r)$ defined in
    Eq.~(\ref{eq:orient}) of the water particles around a solute with
    radius $R=11$\AA~ carrying no charge ($\Sn$, solid line), and
    charge $q=\pm 5$ (${\rm S}_\pm$, dashed lines), and $q=\pm 10$
    (${\rm S}_\pm$, dotted lines).}
\label{fig:orient} 
\end{center}
\end{figure}

\begin{figure}
  \begin{center}
\includegraphics[width=8cm,angle=0.,clip]{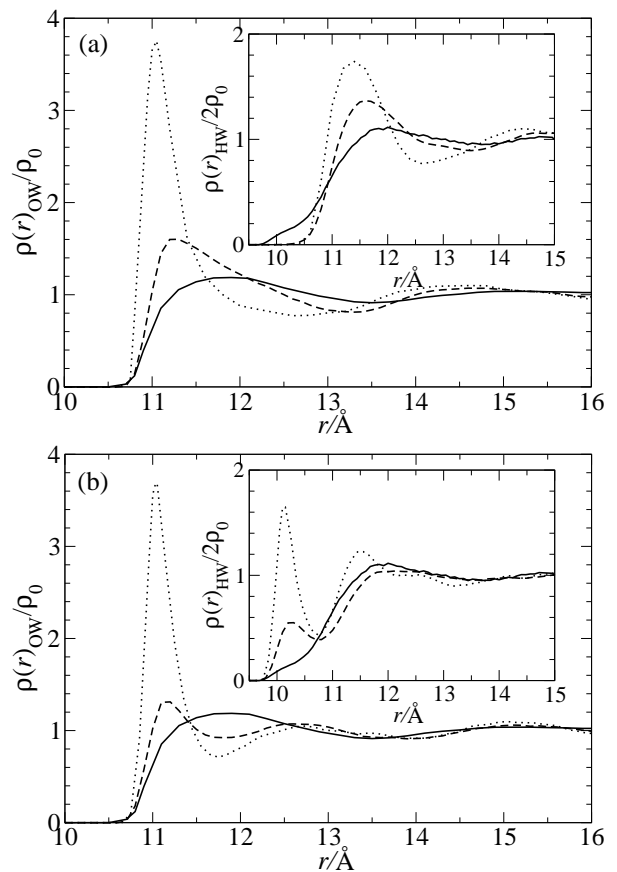}
    \caption{Density profiles of water oxygen and hydrogen atoms
      (inset) around a (a) positively ($\Sp$) and (b) negatively
      ($\Sm$) charged solute with radius $R=11$\AA~ and central charge
      $q=\pm 5$ (dashed line), and $q=\pm 10$ (dotted line). We also
      plot the result for the $\Sn$ model (solid line) with the same
      radius.}
\label{fig:profiles_charged} 
\end{center}
\end{figure}

This observation may be quantified by considering the following
orientational order parameter:
\begin{eqnarray}
P(r)=\left<\frac{\vec\omega\cdot\vec r}{|\vec\omega||\vec r|}\right>_r,
\label{eq:orient}
\end{eqnarray}
where $\vec \omega$ is the water molecule orientation vector
(cf. Fig.~1), and the configurational average is taken for a fixed
distance $r$ from the solute center to the O-site of the water
molecules. MD results for a neutral solute of radius $R=11$\AA~ are
plotted in Fig.~\ref{fig:orient}.  The curve $P(r)$ takes slightly
positive values ($P\simeq 0.05$) for water in the first solvation
shell ($r\simeq 12.5$\AA), indicating a weak tendency of the hydrogen
atoms to point away from the solute.

The effect of charging the solute is illustrated in
Fig.~\ref{fig:profiles_charged}(a) and (b), where the RDFs are plotted
for a fixed radius $R=11$\AA, and charges $q=0,\pm 5$, and $\pm 10$,
within the $\Sn$ and ${\rm S}_\pm$ models (neutral, or uniformly
charged solutes). For positive charges,
Fig.~\ref{fig:profiles_charged}(a), the height of the first peak in
both oxygen and hydrogen RDFs is seen to shift to shorter distances,
and to increase as $q$ increases, signaling an effective attraction of
the dipolar solvent molecules due to the radial electric field. The
trend is similar for negative charges, as regards the oxygen
RDF. However the initial first peak (at $r\simeq 12$\AA~ for $q=0$) in
the hydrogen RDF is seen to split into a prepeak around $r\simeq
10$\AA~ and a broad feature close to the initial peak when
$q=-5$. This points to a reorientation of the water molecules in the
first hydration shell, with the positive hydrogen atom preferring
being closer to the surface of the anionic solute. Further decrease of
the negative charge $(q=-10)$ consolidates this structure, with two
hydrogen peaks growing in amplitude
(cf. Fig.~\ref{fig:profiles_charged}(b)). Interestingly, while the
hydrogen RDFs are very sensitive to the sign of the solute charge
($\Sp$ versus $\Sm$), the amplitudes of the first peaks of the oxygen
RDFs are nearly independent of this sign, but the peak around the
positive solute appears to be broader, signaling a larger water
coordination number in the first solvation shell. The corresponding
orientational order parameter $P(r)$ is plotted in
Fig.~\ref{fig:orient}. The absolute value of $P(r)$ has maxima at
contact $r\simeq 11$\AA~ and approximately one water diameter further
away ($r\simeq 13.5$\AA), and increases with absolute charge,
irrespective of the sign of the charge carried by the solute. Closer
inspection of the curves in Fig.~\ref{fig:orient} reveals, however, a
significant asymmetry, if not in the overall shape of $P(r)$, at least
in the amplitudes, which are typically $20\%$ larger for the negative
solute.  Anion/cation hydration asymmetry had already been reported
for microscopic ions in aqueous solution.\cite{latimer:jcp:1939}

\begin{figure}
  \begin{center}
\includegraphics[width=8cm,angle=0.,clip]{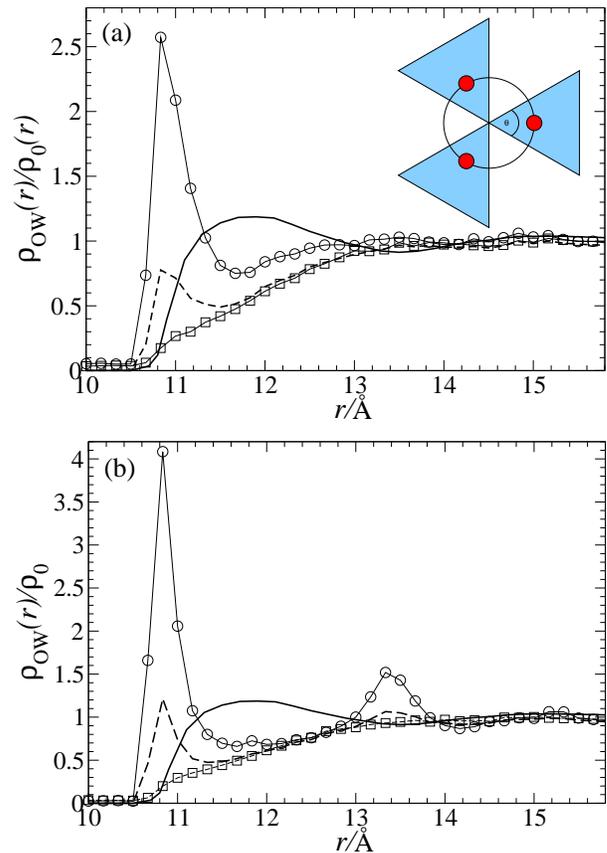}
    \caption{Oxygen density profiles around the tetrahedral (a)
    positive $\Tp$ and (b) negative $\Tm$ solutes. The curves are
    for a full angular average (long dashed lines), cone averages
    around the charges (circles), and averages of water excluded by
    these cones (squares), as explained in section
    \ref{sec:structure}.  We also plot the density profile around a
    neutral solute $\Sn$ (solid line) for
    comparison. The inset in (a) sketches a two dimensional projection
    of a tetrahedral solute. The cone averages are performed over the
    water molecules in the grey cones (opening angle
    $\Theta=30^\circ$) containing a surface charge (dark circle).}

\label{fig:profiles_proteins} 
\end{center}
\end{figure}

\begin{figure}
  \begin{center}
\includegraphics[width=8cm,angle=0.,clip]{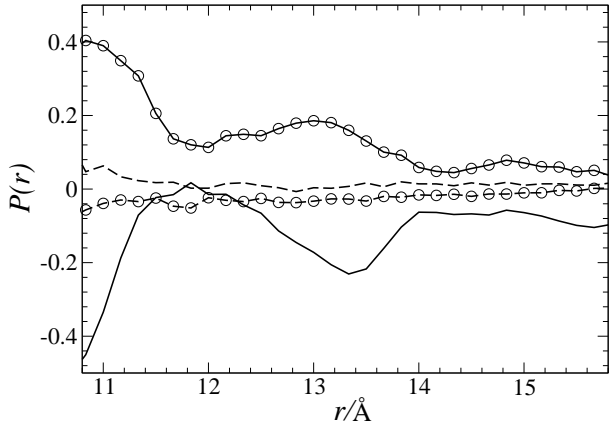}
    \caption{Orientation profiles $P(r)$ around the tetrahedral
    positive $\Tp$ (circles) and negative $\Tm$ (no symbols)
    solutes. Cone averages around charges (solid lines) are compared to
    averages of water molecules excluded by those cones (long dashed
    lines).}
\label{fig:orient_protein} 
\end{center}
\end{figure}

We now turn to the structure of water around a solute with an
inhomogeneous charge distribution, restricting the discussion to the
tetrahedral $\Tp$ and $\Tm$ models. In order to characterize the
anisotropy of the problem, it is desirable to distinguish between
water molecules close to the four surface charges, and the remaining
water surrounding the solute. We achieve this by averaging over water
molecules whose centers fall either inside or outside well-defined
cones whose axes coincide with the radii joining the solute center and
the surface charges and whose vertices coincide with the solute
center. A sketch of the two-dimensional projection of one of the sides
of the tetrahedron and its associated cones is shown in the inset to
Fig.~\ref{fig:profiles_proteins}(a). Averages are taken over cones of
opening angle $\Theta=30^\circ$, high enough to accommodate the first
two solvation shells around a surface charge. The results for the
oxygen density profiles for water molecules inside or outside the
cones are shown in Figs.~\ref{fig:profiles_proteins}(a) and (b) for
the $\Tp$ and $\Tm$ models, respectively. The water molecules inside
the cones exhibit a typical solvation shell structure with a large
first peak, and a much lower second peak, which is hardly visible in
the $\Tp$ case. Outside the four cones, water appears to be highly
depleted compared to its distribution around a neutral $\Sn$ solute,
up to a radial distance $r\simeq 13$\AA. Thus the 25$\%$ of the solute
surface area inside the cones act as hydrophilic ``patches'' while the
remaining 75\% are hydrophobic. Interestingly, if an angular average
is taken over the total solute area, the mean density of water close
to the surface ($\lesssim 13$\AA) of a $\Tp$ or $\Tm$ solute is
significantly smaller than the corresponding density around a neutral
$\Sn$ solute ! Integration of the water density profile up to
$r=13$\AA~ yields coordination numbers (numbers of water molecules) of
133 for $\Sn$, 92 for $\Tp$, and 95 for $\Tm$, showing a 30$\%$
depletion of water around the T-solutes compared to $\Sn$. The
orientational order parameter (3) for the $\Tp$ and $\Tm$ models are
plotted versus $r$ in Fig~\ref{fig:orient_protein}. The orientational
order of water molecules inside the cones is seen to be similar to
that around homogeneously charged solutes $\Sp$ or $\Sm$
(cf.~Fig.~\ref{fig:orient}). In the depleted volumes outside the cones
the water molecules shows little orientational order; if anything they
tend to orient in the direction opposite to the mean orientation
inside the cones.

Qualitatively similar observations hold for the distribution of water
molecules around solutes with cubic charge distribution (models $\Cp$
or $\Cm$), but obviously the volumes depleted of water are now
smaller, since the ``hydrophobic patches'' have shrunk now to only
half of the solute surface area. The MD simulations for the solutes
with non-vanishing net charge were carried out with explicit
counterions. Test runs where these counterions were replaced by a
uniform neutralizing background showed no differences within the
statistical uncertainties.

\subsection{Solvation free energy}
\label{sec:solvation}

The solvation free energy is equal to the reversible work required for
transferring a solute from vacuum into a solvent. For neutral solutes
in water, the solvation free energy is generally
positive,\cite{hummer:jpc:1996,lum:jpc} and for atomic-size solutes,
it stems mainly from the entropy cost of the restructuring water
molecules around the solute. For larger spherical solutes, a cross
over in the variation of the solvation free energy with radius occurs
typically around 1 nm.\cite{lum:jpc} The classic Born model
\cite{born} provides the simplest approach to the solvation free
energy of charged spherical solutes. The solvent is treated as a
dielectric continuum of permittivity $\epsilon$ and the hydration free
energy increases quadratically with solute charge and is proportional
to the inverse of the Born radius $R_{\rm B}$, according to
\begin{eqnarray}
  \Delta\mu_{\rm B}=-\frac{q^2e^2}{8\pi\epsilon_0 R_{\rm B}}(1-1/\epsilon).
\label{eq:born}
\end{eqnarray}
Note that the solvation free energy is a difference in chemical
potential of the solute as it is moved from vacuum into the
solvent. Hydration free energies from the Born model agree well with
experimental values, once the unknown parameter $R_{\rm B}$ is
defined. The Born radius for an ion can deviate substantially from its
Pauling radius (a measure of the size of an
ion).\cite{latimer:jcp:1939}

We have obtained solvation free energies for our model solutes by
thermodynamic integration, using the general formula
\begin{eqnarray}
\Delta\mu_{\rm sim}=\int_{\lambda_{0}}^{\lambda_{1}} d\lambda \left<\frac{\partial V_{\rm sol}}{\partial \lambda}\right>_{\lambda},
\label{eq:ti}
\end{eqnarray}
where the coupling parameter $\lambda$ gradually ``switches on'' the
interaction (2) between the solute and the solvent from an initial
state $\lambda=\lambda_0$ (say a neutral point solute) to a final
state $\lambda=\lambda_1$ corresponding to the complete
solute/solvent system.  The brackets $<..>_\lambda$ denote a
statistical average over all solute-solvent configurations for a
solute-solvent coupling characterized by $V_{\rm sol}(\lambda)$. The
index sim in $\Delta\mu_{\rm sim}$ indicates that the estimate of the
solvation free energy is based on MD simulations of a finite sample;
finite size corrections will be added as explained later.

\begin{figure}
  \begin{center}
\includegraphics[width=8cm,angle=0.,clip]{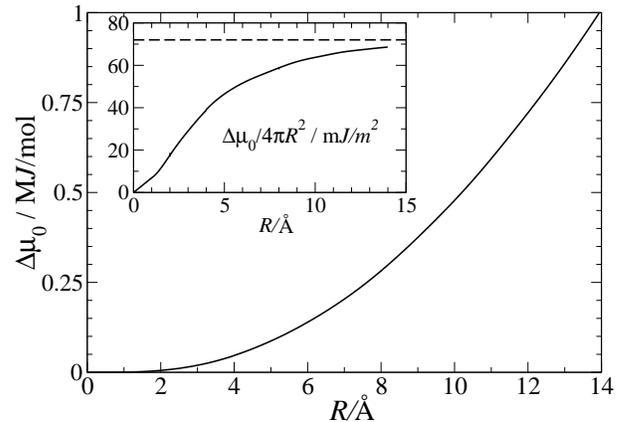}
    \caption{Solvation free energy $\Delta\mu_0$ of a neutral
    spherical solute $\Sn$ in water. The inset shows $\Delta\mu_0$ divided
    by the sphere surface showing an asymptotic approach for large $R$
    to the liquid-vapor surface tension of water (dashed line),
    $\gamma\approx 72$mJ/${\rm m^2}$.}
\label{fig:freee_neutral} 
\end{center}
\end{figure}

In practice, we proceded in two steps. In a first stage, we computed
the solvation free energy $\Delta\mu_0$ of a neutral spherical solute
(model $\Sn$), as a function of its radius $R$. The second step is to
charge up the initially neutral solute to the final charge pattern. In
step one, the coupling parameter $\lambda$ in Eq.~(\ref{eq:ti}) is simply the
radius itself, and $\Delta\mu_0$ is consequently the work required to
blow up the solute against the normal force exerted by the solvent,
integrated over the particle surface; in this case the force is just
the radial derivative of the first term on the rhs of eq
(\ref{eq:totalsolutepot}). This is implemented, in practice, by
starting from $\lambda_0=R=0$, and increasing the radius by steps of
$\Delta\lambda=\Delta R=1$\AA, up to $\lambda_1=14$\AA~ (the largest
neutral solute considered in the present work). The averaged radial
force, as obtained from the MD simulations for various $R$, is
interpolated with a cubic spline and integrated to yield
$\Delta\mu_0$. The resulting solvation free energy is plotted in
Fig.~\ref{fig:freee_neutral}, and is seen to increase monotonically
with $R$. The solvation free energy per unit area, $\Delta\mu_0/4\pi
R^2$, is plotted in the inset to Fig.~\ref{fig:freee_neutral}, and is
seen to approach asymptotically a constant value for radii $R\gtrsim
10$\AA; the latter is close to the liquid-vapor surface tension of
water, $\gamma=72$mJ/m$^2$. This behavior is close to that reported by
Lum {\it et al.}\cite{lum:jpc} for hard sphere cavities in water. The
``softer'' solute-solvent pair potential (1) used in the present work
does not modify the solvation process significantly, compared to the
case of hard sphere solutes.

\begin{figure}
  \begin{center}
\includegraphics[width=8cm,angle=0.,clip]{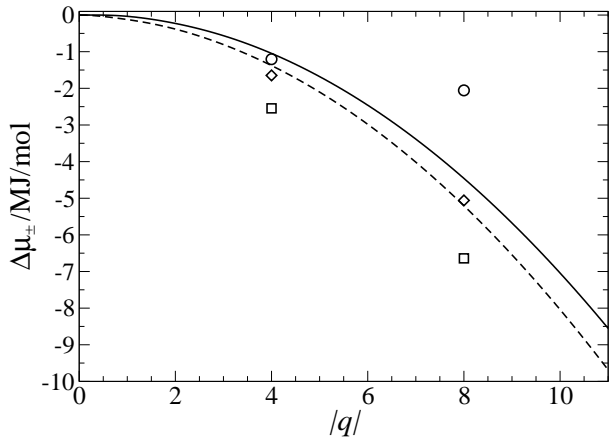}
    \caption{Excess solvation free energy $\Delta\mu_\pm$ of charging
    a spherical solute of radius $R=11$\AA~ in water homogeneously to
    a charge $q$ (solid line) and $-q$ (long dashed line). The symbols
    denote the solvation free energies of charging the ${\rm
    T}_{0\pm}$ and ${\rm C}_{0\pm}$ models in water. The symbols at
    $q=4$ are for the tetrahedron, at $q=8$ for the cube. The overall
    charge of the ${\rm T}_{0\pm}$ and ${\rm C}_{0\pm}$ solutes is
    zero (circles), positive (diamonds), and negative (squares).}
\label{fig:freee_charged} 
\end{center}
\end{figure}

In the second stage, to go from the neutral to the charged solute, the
charges of the $N_c$ sites on the solute are gradually turned on,
i.e. $q_\alpha(\lambda)=\lambda q_\alpha$ $(1\leq \alpha \leq N_c)$,
where $\lambda$ is varied from 0 to 1. The quantity to be averaged in
Eq.~(\ref{eq:ti}) is now the total electrostatic energy of the solute in
the field of the water molecules and their periodic images; the
statistical average is to be taken over all Boltzmann-weighted water
configurations when the electrostatic solute/solvent coupling is
multiplied by $\lambda$. Since for any $\lambda>0$ the system carries
a net charge, a compensating uniform background charge must be
included in evaluating the Coulombic part of
Eq.~(\ref{eq:totalsolutepot}) by Ewald summation. If the self
interaction energy of the solute with its own images and the
neutralizing background is properly included, the resulting free
energies are virtually independent of the size $L$ of the simulation
box,\cite{hummer:jpc:1996, hummer:jcp:1997} see the Appendix
B. Results for $\Delta\mu_{\pm}$ corresponding to a uniformly charged
solute with radius $R=11$\AA~ are plotted in
Fig.~\ref{fig:freee_charged} as a function of $q$, for anionic and
cationic solutes. Note that this is the excess free energy for
charging an initially neutral solute of $R=11$\AA, previously inserted
into the solvent, to a charge $q$. In order to obtain the total
solvation free energy, the contribution $\Delta\mu_0$ corresponding to
the insertion of the neutral solute in water ($\approx 0.6$MJ/mol for
$R=11$\AA) must be added to $\Delta\mu_\pm$.  The curves are
essentially quadratic on the scale shown, in agreement with Born
theory. Least squares fits of the data to the Born formula (4) yield
$R_{\rm B}^+=9.8$\AA~ and $R_{\rm B}^-=8.6$\AA~, both smaller than the
effective solute radius $R=11$\AA.  The solvation free energy of
cationic solutes is slightly positive for $0<q\lesssim 1$. Such a
behavior has already been reported for small cationic solutes (e.g
Na$^+$)\cite{hummer:jpc:1996,lynden-bell} and may be understood from
the competition between the free energy cost of the rearrangement of
water around the solute, and the electrostatic energy gain of the
dipolar solvent in the electric field of the solute. The latter
contribution appears to dominate already for small $|q|$ in the case
of anionic solutes, for which $\Delta\mu_-$ is always negative. Over
the whole range of absolute charge $|q|$, $\Delta\mu_-$ is
systematically lower than $\Delta\mu_+$, pointing to a preferential
solvation of anionic solutes, again in agreement with earlier findings
for other charged solute models.\cite{hummer:jpc:1996,lynden-bell,garde:2004}
In Sec. \ref{sec:structure} we learned that the restructuring of water
is stronger around an $\Sm$ solute than around its $\Sp$ counterpart,
so that one would expect a higher cost in entropy. Apparently the
closer approach of the hydrogen atoms to the negative solute decreases
the electrostatic contribution to the free energy more than the
positive entropic cost, resulting in an overall lower solvation free
energy.

Solvation free energies for solutes carrying discrete tetrahedral or
cubic charge distributions, corresponding to models $\Tp$, $\Tm$, and
$\Tn$, and $\Cp$, $\Cm$, and $\Cn$ are also shown in
Fig.~\ref{fig:freee_charged}. The contribution to the solvation free
energy from the steric LJ-part of the surface charge interaction with
the water molecules is insignificant and was neglected in the calculation
of $\Delta\mu_\pm$. As in the case of the uniformly charged
solutes, the negative solutes are preferentially solvated compared to
their positive counterparts. The solvation energies of the overall
neutral solutes $\Tn$ and $\Cn$ lie well above those of the charged
solutes (${\rm T}_\pm$ or ${\rm C}_\pm$). In particular $|\Delta\mu|$ of the
overall neutral solute with a cubic charge pattern $\Cn$ is roughly
three times smaller than the corresponding $|\Delta\mu_\pm|$ of the
globally charged solutes $\Cp$ and $\Cm$. This may be a consequence of
the considerable reorganization of water around the solutes with
surface charges of alternating sign, resulting in a substantial cost
in free energy. Finally, the solvation free energies of solutes with
discrete charge patterns (${\rm T}_\pm$ or ${\rm C}_\pm$) are seen to lie
10-20$\%$ below the solvation free energies of their uniformly charged
counterparts ($S_\pm$ with $q=\pm 4$ and $\pm8$).

\section{Effective interaction between two solutes}
\label{sec:force:theory}
We now turn to the main objective of this paper, namely the
determination of the effective, solvent-mediated interaction between
two nanometer-sized neutral or charged solutes in water. In subsection
IV A and IV B we derive this interaction from simple macroscopic
considerations, while the MD methodology is presented in IV C.
\subsection{Phenomenological theory for charged plates}
A simple macroscopic argument, similar to Kelvin's theory of capillary
condensation predicts that water near liquid/vapor coexistence will
undergo ``drying'' when confined between two hydrophilic plates, below
a critical distance $D_c$ separating these plates.\cite{lum:jpc} We
extended the argument to the case of charged
plates,\cite{dzubiella:jcp:2003} showing that $D_c$ is strongly
reduced by the electrostatic energy associated with the surface charge
carried by the plates. The macroscopic argument is further refined
hereafter.  Consider two parallel plate-like solutes of area $A_1$,
separated by a distance $D$ and carrying opposite surface charges
$\pm\sigma$, immersed in a polar solvent of dielectric permittivity
$\epsilon$.  Neglecting edge effects (an approximation valid as long
as $D\ll A_1^{1/2}$, the electric field between the plates is
$E_{0}/\epsilon$ with $E_{0}=\sigma/\epsilon_{0}$. We require the
difference in the grand potential between the situations where the
liquid solvent ($l$) or its vapor ($g$) fill the volume $A_1 D$
between the two plates:
\begin{eqnarray}
\Omega_{\alpha}  =  -P_{\alpha} A_1D  + \frac{1}{2}\epsilon_{0}\frac{E^{2}_{0}}{\epsilon_{\alpha}}A_1D 
 +  2\gamma_{w\alpha}A_1 +\gamma_{l\alpha}A_2,
\label{eq:plates1}
\end{eqnarray}
where $P_{\alpha}$ is the pressure of phase $\alpha=l,g$ and
$\gamma_{w\alpha}$ the surface tension between phase $\alpha$ and the
plate (``wall"). $A_2$ is the area of the liquid-vapor interface
limited by the edges of the two opposite plates, which is created when
the volume between the plates is filled by vapor. $\gamma_{l\alpha}$
is the liquid-vapor surface tension when $\alpha=g$, and vanishes of
course when $\alpha=l$. The last term in Eq.~(\ref{eq:plates1}) may be
neglected for infinitely large plates.\cite{dzubiella:jcp:2003}
Consider a state close to phase coexistence at temperature $T$, and
let $\delta\mu=\mu-\mu_{\rm sat}$ be the positive deviation of the
chemical potential from its saturation value. Expanding the
$P_{\alpha}$ to linear order around their common value at saturation,
one arrives at the following expression for the difference in grand
potentials:
\begin{eqnarray}
\Omega_{g}-\Omega_{l}&=&(\rho_{ l}-\rho_{
  g})\delta\mu A_1D+\frac{\epsilon_{0}}{2}E^{2}_{0}(\frac{1}{\epsilon_{
  g}}-\frac{1}{\epsilon_{l}})A_1D\nonumber\\&+& 2(\gamma_{ w g}-\gamma_{w
  l})A_1-\gamma_{lg}A_2.
\end{eqnarray}
In water $\epsilon_{l}\equiv\epsilon\gg\epsilon_{g}\simeq 1$,
$\rho_{g}\ll\rho_{l}$ (except near critical conditions) and
$\gamma_{wl}-\gamma_{wg}=\gamma_{lg}\equiv\gamma$ for a purely
hydrophobic surface. Moreover $A_2=UD$, where $U$ is the
circumference of one plate. Hence:
\begin{eqnarray}
\Delta\Omega =\left(\rho_{l}\delta\mu+\frac{\epsilon_{0}}{2}E^{2}_{0}\right)A_1D- 2\gamma A_1 +\gamma UD.
\label{eq:deltaomega}
\end{eqnarray}
$\Delta\Omega=\Omega_{g}-\Omega_{l}$ is the reversible work bringing
the two plates from infinite separation (when the volume between them
is filled by liquid) to a distance $D$, at which ``drying'' has already
occurred. At contact, $\Delta\Omega(D=0)=-2\gamma A_1$, and
$\Delta\Omega$ increases linearly with $D$. The range of the purely
attractive potential is defined by the distance $D_c$ at which
$\Delta\Omega=0$; for $D>D_c$, the liquid is the preferred phase between the plates; for $D<D_c$, ``drying'' occurs.
From Eq.~(\ref{eq:deltaomega}) we obtain
\begin{eqnarray} 
D_{\rm c}\simeq\frac{2\gamma}{\rho_{l}\delta\mu+\frac{\epsilon_{0}}{2}E^{2}_{0}+\gamma U/A_1}.
\label{eq:Dc}
\end{eqnarray}
Near the liquid-vapor transition of water $\delta\mu\ll k_{B}T$, and
may be neglected compared to the surface tension term in the
denominator. Consider circular plates of radius $R$; then $U/A_1=2R$,
and if they are uncharged ($E_0=0$), $D_c=R$, which is in agreemeent
with previous calculations of the mean force between plate-like
solutes in water \cite{wallquist:jpc} and in a LJ-fluid.\cite{bolhuis}
In the case of high
surface charges ($\sigma\lesssim e/{\rm nm}^{2}$), $E_0$ can be as
large as $10^{10}$V/m, and the corresponding electrostatic term in the
denominator becomes comparable to the surface term for solute sizes of
a few nm. This leads to a strong reduction of $D_{\rm c}$ compared to
the case of neutral solutes. This reduction of $D_{\rm c}$ hints at a
considerable weakening of the hydrophobic interaction between two
solutes when the latter are charged. This trend will be confirmed by
the MD results in Sec.~V. Note however that the simple macroscopic
model ignores molecular details, and that its prediction does not, a
priori, apply to solutes carrying discrete charge patterns
(i.e. ``hydrophilic patches'') for which a more microscopic
description is required.

\subsection{Phenomenological theory for spherical solutes}
\label{sec:theory}
\begin{figure}
  \begin{center}
\includegraphics[width=8cm,angle=0.,clip]{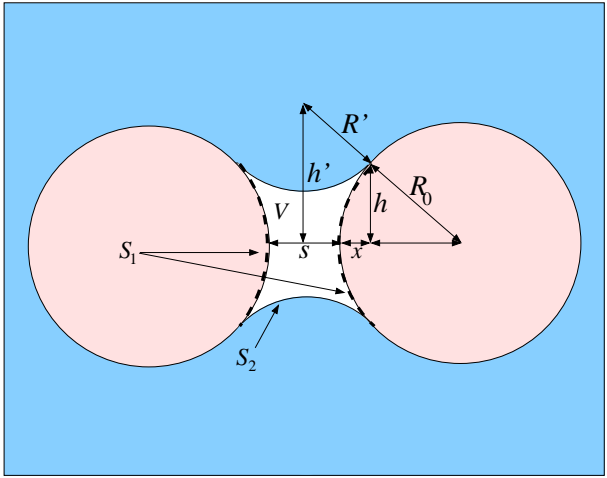}
    \caption{Sketch of two spherical solutes of radius $R_0$ at a
    surface-to-surface distance $s$. The white volume $V$ between the
    solutes approximates the region depleted of water. $S_1$ and $S_2$ are
    the surrounding solute-vapor and liquid-vapor surface areas,
    resp. $R'$ is the radius of the curved surface $S_2$.}
\label{fig:sketch} 
\end{center}
\end{figure}

The previous model can be extended to the case of neutral or charged
spherical solutes with radius $R_0$ as follows (see
Fig.~\ref{fig:sketch}). 
When ``drying'' occurs, the simulation data
(cf. Fig.~\ref{fig:contour_neutral} (b) or (c)) suggest that a
cylindrically symmetric domain bounded by the two spherical solute
surfaces ($S_1$) and the curved liquid-vapor meniscus ($S_2$) is
filled with vapor. The surface $S_2$ is assumed to touch the solute
spheres tangentially (contact angle $\pi$) and to have a radius of
curvature $R'$. For a given surface-to-surface distance $s$ of the two
solutes, the volume $V$ of the ``dry'' domain and the areas $S_1$ and
$S_2$ are conveniently expressed in terms of the single parameter $x$,
as depicted in Fig.~\ref{fig:sketch}. If $x=0$, the vapor domain
shrinks to zero, i.e. the space between the solutes is filled with
liquid, while for $x=R_0$ the vapor occupies a cylindrical volume
$V=\pi R_0^2[(2R_0+s)-4R_0/3]$. For intermediate values of $x$, the
areas $S_1$ and $S_2$ are given by
\begin{eqnarray}
 S_1  (x)  &=&  4\pi R_0x \nonumber \\
 S_2  (x)  &=&  4\pi R'\left[h' \arcsin\left(\frac{x+s/2}{R'}\right)-(x+s/2)\right] \nonumber \\
 R' &  =  & R_0(x+s/2)/(R_0-x) \nonumber \\ h'  & = & \frac{R_0+R'}{R_0}\sqrt{2R_0x-x^2}.
\end{eqnarray}
\begin{figure}
  \begin{center}
\includegraphics[width=8cm,angle=0.,clip]{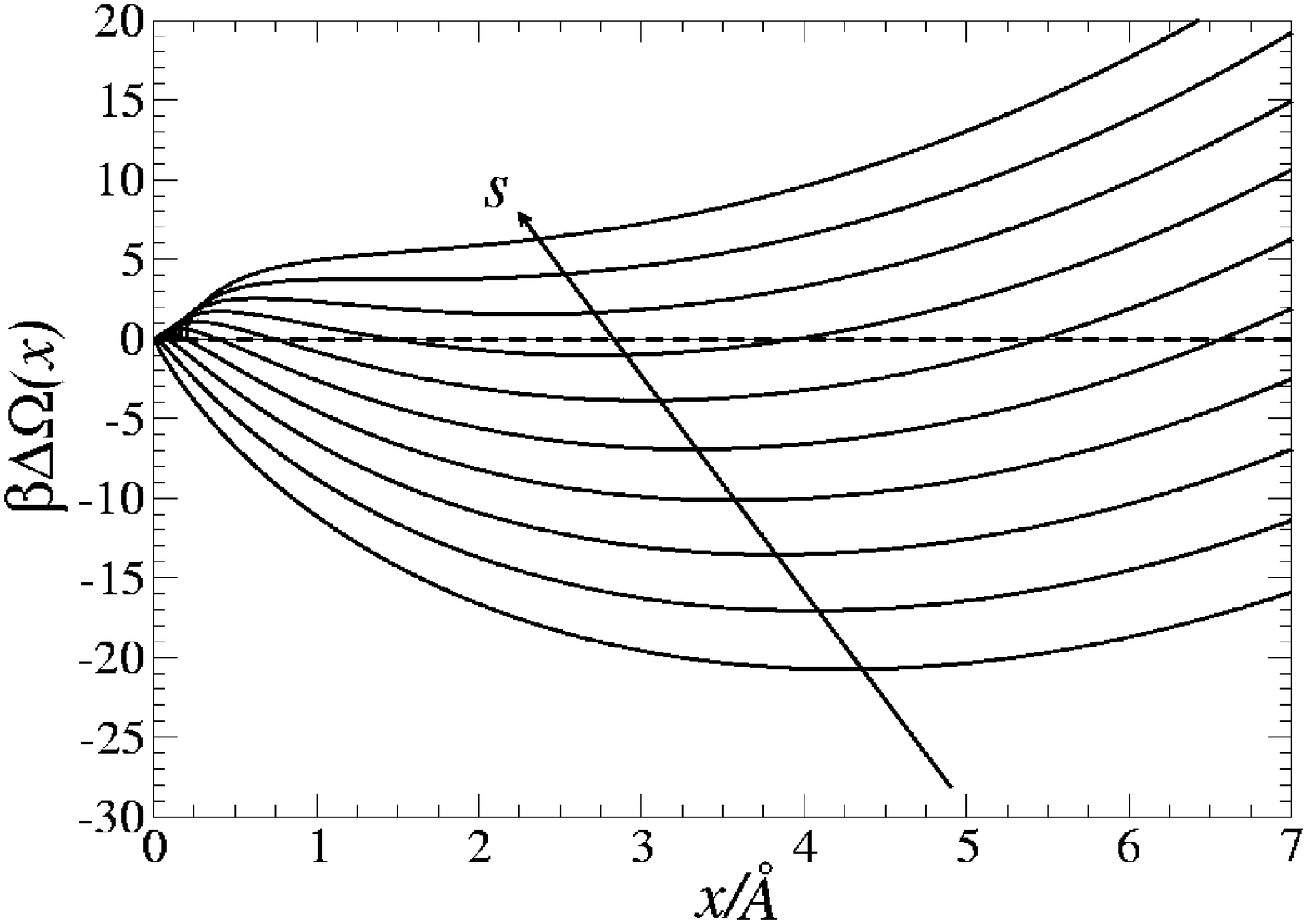}
    \caption{Free energy $\Delta\Omega$ of dried states between
    neutral spherical solutes with radius $R_0=10$ according to
    Eq.~(\ref{eq:domega}). The volume of the empty state is described
    by the parameter $x$ for a given geometry, see
    Fig.~\ref{fig:sketch}. The most bottom curve is for a
    surface-to-surface distance $s=0$, then $s$ is incremented by
    0.5\AA~ steps.}
\label{fig:omegax} 
\end{center}
\end{figure}
The difference in grand potential between the ``dry'' and filled
states is then given (in absence of the electric charges) by the
following generalization of Eq.~(\ref{eq:deltaomega}):
\begin{eqnarray}
\Delta\Omega=\rho_l \delta\mu V(x)-\gamma_{\rm s} S_1(x)+\gamma S_2(x),
\label{eq:domega}
\end{eqnarray}
where $\gamma_s$ is the surface tension of the solute-liquid interface
and $\gamma$ the liquid-vapor surface tension.  The first term in
Eq.~(\ref{eq:domega}) is the bulk free energy for creating a cavity of
volume $V(x)$ in water, and favors the filled state. Again, near
liquid-vapor equilibrium $\delta\mu\ll k_{B}T$, and the volume term
may be safely neglected. The second and third terms in
Eq.~(\ref{eq:domega}) are the surface free energies for decreasing the
solute-liquid and increasing the liquid-vapor interface,
respectively. For simplicity we first assume
$\gamma=\gamma_s$\cite{huang:jpc}; effects arising from
$\gamma\neq\gamma_s$ will be discussed later. In the following we use
the liquid-vapor surface tension of water $\gamma=0.174
k_BT{\rm\AA}^{-2}$. $\Delta\Omega(x)$ is plotted versus the geometric
control parameter $x$ in Fig.~\ref{fig:omegax} for a solute of radius
$R_0=1$nm, and various surface-to-surface distances $s$. At contact
($s=0$), $\Delta\Omega(x)$ exhibits a single negative minimum at the
non-zero value $x=x_{\rm min}(s=0)$ signaling that the ``dry'' state
with volume $V(x_{\rm min})$ is stable. As $s$ increases, the global
minimum is raised to less negative energy values and occurs at smaller
values of $x$, while a second local minimum appears at $x=0$,
corresponding to a metastable, filled state. At the critical value
$s_c\simeq 3.2$\AA, the global minimum jumps from the non zero value
of $x$ to $x=x_{\rm min}(s_c)=0$. For $s>s_c$, the filled state is
favored. The effective interaction $w(s)$ between two solutes is equal
to the free energy difference in bringing them from infinite
separation to a surface-to-surface distance $s$, and is given by
Eq.~(\ref{eq:domega}), i.e. $w(s) =\Delta\Omega(s;x_{\rm
min}(s))$. The energy at the global minimum is negative for all
$s<s_c$ ($x_{\rm min}>0$), so that the interaction is always
attractive. $s_c$ is the range of the interaction; at $s=s_c$, $w
(s)=\Delta\Omega=0$, and $\gamma_{\rm s} S_1=\gamma S_2$. 

\begin{figure}
  \begin{center}
\includegraphics[width=8cm,angle=0.,clip]{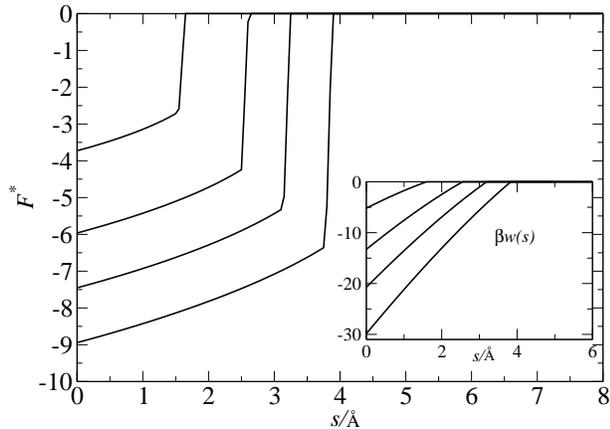}
    \caption{Results from the phenomenological theory in section \ref{sec:theory}
      for the mean force, $F^{*}=\beta F$\AA, between two neutral
      ($q=0$) spheres in SPC/E water. Results are plotted for solute
      radii $R_0=5$\AA~, $R_0=8$\AA~, $R_0=10$\AA~, and
      $R_0=12$\AA~. The inset shows the integrated force (potential of
      mean force). The depth and range of the force and potential
      increases with $R$.}
\label{fig:forces_theory} 
\end{center}
\end{figure}

The effective potentials and forces between two solutes of different
radii $R_0=5,8,10,12$\AA~ from Eq.~(\ref{eq:domega}) are plotted
in Fig.~\ref{fig:forces_theory}. A detailed numerical investigation
shows that, assuming $\gamma_s=\gamma$, the range of the potential
(and of the resulting effective force) is $s_c\simeq 0.32R_0$, scaling
linearly with solute radius, independently of $\gamma$. The contact
value of the potential is $w(0) \simeq -1.19\gamma R_0^2$, scaling
with the solute surface area.  The contact value of the force is $
F(0)\simeq -4.28\gamma R_0$. The force increases roughly linearly with
a slope independent of $R_0$ (cf. Fig.~12). The results are accurately
represented by
\begin{eqnarray}
 F(s) =\gamma
\begin{cases}
-a_1 R_0 + a_2s & {\text {for}} \,\,s \leq s_c=a_3R_0 ; \cr
   0 & {\rm otherwise}. \cr
\end{cases}
\label{eq:force}
\end{eqnarray}
and
\begin{eqnarray}
 w(s) =\gamma
\begin{cases}
-a_0 R_0^2 +a_1R_0s-\frac{a_2}{2}s^2 & {\text {for}} \,\,s \leq s_c ; \cr
   0 & {\rm otherwise}. \cr
\end{cases}
\label{eq:pot}
\end{eqnarray}
with $a_0=1.19$, $a_1=4.28$, $a_3=0.32$, and $s_c=a_3 R_0$.
If the force is assumed to be linear in $s$, $a_2$ is fixed by
the other constants according to $a_2=2(a_1a_3-a_0)/a_3^2\simeq 3.51$.
The linear scaling of the contact value of the force with solute
radius $R$ may be rationalized by a simple consideration of the
potential of mean force for plates ($R=\infty$) at contact,
$w(0)=-2\gamma$ and an application of the Derjaguin approximation,
valid for weakly curved substrates (i.e. large $R$); \cite{louis:jcp}
this leads to the estimate $F(s=0)=-2\pi R\gamma$, which indeed
predicts linear scaling.

Near a ``realistic'' solute, water will have a surface tension
$\gamma_s\neq\gamma$, due to Van-der-Waals and electrostatic
interactions, as well as the influence of curvature for small
solutes. One may expect $\gamma_s < \gamma$, and lowering $\gamma_s$
will lead to less hydrophobic attraction.  A possible
approach to include the electrostatic field effects due to a net
charge on the solutes is to absorb these effects into the
solute-liquid surface tension.  A naive procedure is to approximate
$\gamma_s$ by the solvation free energy per unit area of a charged
solute. In section \ref{sec:solvation} it was shown that the Born
expression (4) fits the MD data for uniformly charged solutes quite
well, once the Born radius $R_B$ has been adjusted.  Thus one may
write for large solutes $R\gtrsim1$nm (neglecting $1/\epsilon$ compared to 1):
\begin{eqnarray}
\gamma_s=\gamma-\frac{q^2e^2}{32\pi^2\epsilon_0 R_B^3}, 
\label{eq:gamma_charged}
\end{eqnarray}
which indeed lowers the hydrophobic attraction when charge is added to
the solutes. No hydrophobic attraction occurs when $\gamma_s=0$, so
that the corresponding critical charge $q_c$ satisfies:
\begin{eqnarray}
{q_c^2}={32\pi^2 e^2\epsilon_0 \gamma R_B^3}
\label{eq:gamma_charged}
\end{eqnarray}
For $R_B \approx 1$nm, one obtains $|q_c|\approx 3$, which in view of the
sensitivity to $R_B$, yields at least  the right order of magnitude, since 
the MD data discussed later suggest $|q_c|\approx 8$.

\subsection{Forces and potentials from simulation}
In the MD simulations, the mean force between two solutes was
calculated by placing them at fixed positions ${{\vec R}_1}$ and
${{\vec R}_2}$ along the body diagonal of the simulation cell, and
averaging over water configurations generated during the runs which
extended typically over 1-3 ns. The averaging was performed as long
the statistical error was larger than $\Delta \beta F{\rm \AA}=0.5$,
approximately twice the symbol size in the figures showing the
forces. Separate MD simulations have to be carried out for each
center-to-center distance ${{\vec R}_{12}}={{\vec R}_1}-{{\vec R}_2}$,
i.e. for a series of surface-to-surface distances $s=R_{12}-2R_0$.
The force acting on solute 1 is estimated from the statistical average
of the gradient of the total interaction energy $V_{\rm sol}$ in
Eq.~(\ref{eq:totalsolutepot}):
\begin{equation}
{\vec F}_1(R_{12}) = \left\langle -\nabla V_{\rm sol}(R_{12})\right\rangle_{{\vec R}_{12}},
\label{f1.eq}
\end{equation}
where the constrained statistical average is taken over solvent
configurations, when the two solutes are held fixed at a separation
$\vec R_{12}$. Note that while the solute translational degrees are
frozen, they rotate freely under the action of the torques exerted by
the solvent and the other solute. In other words, the calculated
effective forces are orientationally averaged. By symmetry, $\vec
F_2(\vec R_{12})=-\vec F_1(\vec R_{12})$. The magnitude of the
effective force is obtained by projecting onto the vector $\vec
R_{12}$:
\begin{equation}
F(R_{12}) = \frac{{\vec R}_1 - {\vec R}_2}{R_{12}}\cdot
                      {\vec F}_1(R_{12}).
\label{fdep.eq}
\end{equation}
The resulting effective solute-solute potential follows from:
\begin{equation}
w(R_{12})=\int_{R_{12}}^\infty F(R)dR.
\label{eq:int}
\end{equation}
In simulations where counterions are present, the sum of all pair
interactions between the latter and the solute must be added to
$V_{\rm sol}$ in Eq.~(\ref{f1.eq}), i.e. the mean force acting on the
solute is the statistical average of the sum of the instantaneous
forces exerted by all water molecules and ions on the solute, in the
presence of a fixed second solute.

\section{Molecular dynamics results for profiles and forces}
\label{sec:force:sim}
\subsection{Neutral solutes}

\begin{figure}
  \begin{center}
\includegraphics[width=6cm,angle=0.,clip]{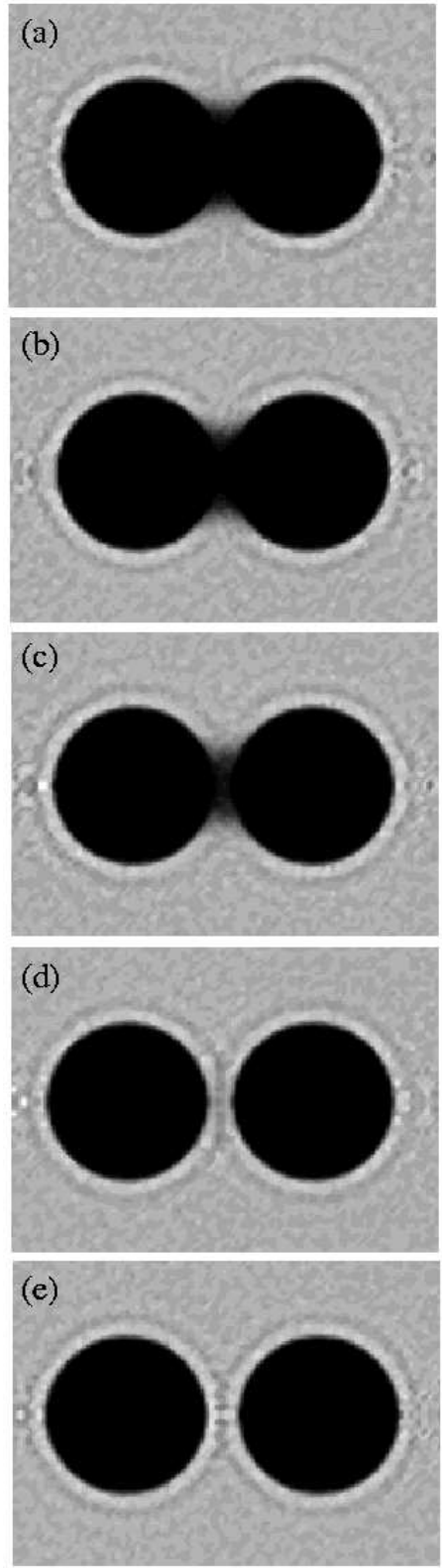}
\caption{Contour density profiles of water around two neutral
$\Sn$ solutes with radius $R=11$\AA~ for surface-to-surface
distances (a) $s=2$\AA, (b) $s=4$\AA, (c) $s=5$\AA, (d) $s=6$\AA,
and (e) $s=7$\AA. The dark and light areas correspond to low and high
densities of water molecules.}
\label{fig:contour_neutral}
\end{center}
\end{figure}

We first consider the case of uncharged solutes. The water density
profiles are illustrated in Figs.~\ref{fig:contour_neutral}(a)-(e) for
the case of solutes of radius $R=11$\AA~ and different
surface-to-surface distances along the $z$-axis joining the
centers. We plot density contours coded by variable shades of
gray. The profiles are calculated using a cylindrical average around the
symmetry axis (center-to-center line).  The profiles show a
considerable depletion (dark region) of the solvent between the
spheres, reminiscent of the observations of Wallquist and Berne for
flatter solutes. \cite{wallquist:jpc} As the surface-to-surface
distance $s$ is increased for fixed radius $R$, the water molecules
penetrate into the region between opposite solute surfaces, as
signaled by a decreasing radius of the dark  region between
the spheres. Eventually at a distance between $s\approx5$\AA~ and
6\AA~ (Figs.~\ref{fig:contour_neutral}(c) and (d)) the region between
the solutes fills with liquid.  When $s\gtrsim 7$\AA, the solvent
layers around an isolated solute are hardly disturbed by the presence
of the other solute.

\begin{figure}
  \begin{center}
\includegraphics[width=8cm,angle=0.,clip]{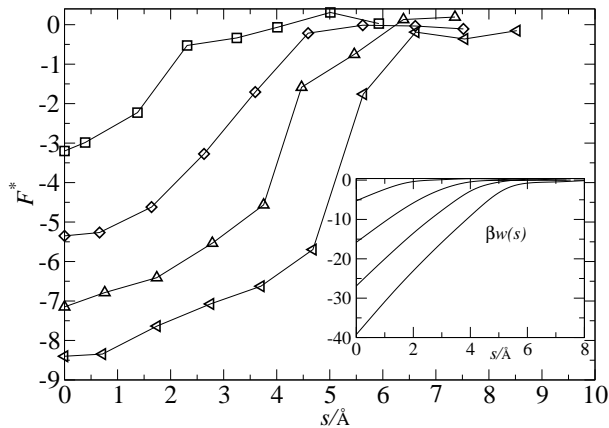}
    \caption{Simulation results (symbols) of the mean force,
      $F^{*}=\beta F$\AA, between
      two neutral $\Sn$ solutes in SPC/E water. The
      lines are guides to the eye. Results are plotted for solute
      radii  $R=6$\AA~ (squares), $R=9$\AA~
      (diamonds), $R=11$\AA~ (triangles pointing up), and $R=13$\AA~ (triangles
      pointing left). The inset shows the integrated force (potential of mean
      force) in obvious order.}  
\label{fig:forces_sim} 
\end{center}
\end{figure}

Examples for the mean force for several radii 6\AA~$\leq R \leq
13$\AA~ are shown in Fig.~\ref{fig:forces_sim}. The largest radii are
of the order of the size of small globular proteins or of oil-in-water
micelles.  The average force obviously goes to zero at large distances
$s$ and for symmetry reasons, it is directed along the
center-to-center axis. As expected from a depletion mechanism, the
force is attractive and its contact values and range increase with
$R$. We observe for all radii that the range of the force is similar
to the distance at which the drying between the solutes vanishes.  The
potentials of mean force $w(s)$ may be calculated for each $R$
according to Eq.~(\ref{eq:int}). The resulting potentials are shown in the inset
to Fig.~\ref{fig:forces_sim}. They closely resemble results obtained
for polymer-induced depletion potentials between spherical colloids,
albeit on different length and energy scales.\cite{louis:jcp} The
force at contact, $F(s=0)$, and the range of the force $s_c$ scales
roughly with $R$ in agreement with the macroscopic model prediction
(14). From the MD data we extract a rough scaling $F(s=0)\approx
3.9\gamma R$ and range $s_c\approx 0.4 R$, reasonably close to the
theoretical estimates from Sec.~IV.B. Overall there is a striking
qualitative and even semi-quantitative agreement between the MD forces
in Fig.~14 and the predictions of tbe macroscopic model in Fig. 12.

\subsection{Uniformly charged solutes}

\begin{figure}
  \begin{center}
\includegraphics[width=8.6cm]{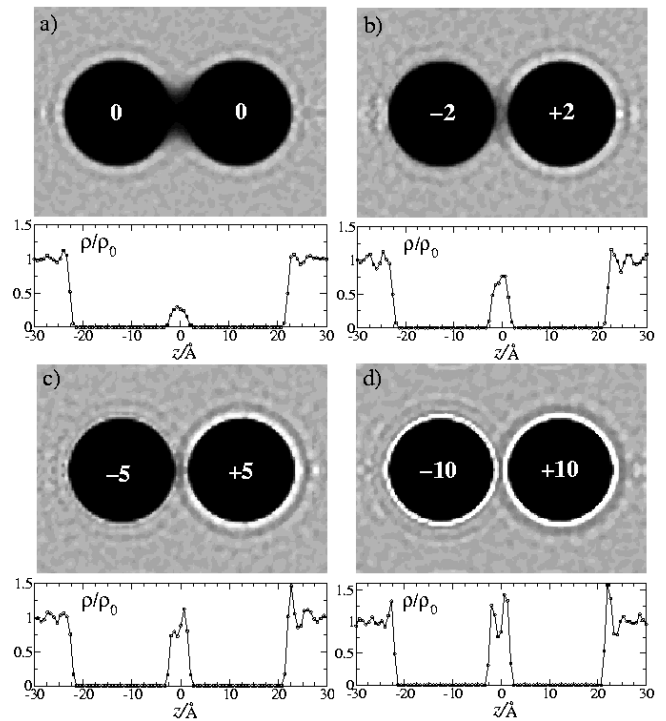}
    \caption{Density profiles of the water molecules around (a) two
      neutral $\Sn$ solutes of radius $R=11$\AA~  and two oppositely
      charged ${\rm S}_\pm$ solutes of radius $R=11$\AA~ carrying a
      charge $\pm qe$ with (b) $|q|=2$, (c) $|q|=5$, and (d)
      $|q|=10$. The surface-to-surface distance in all cases is
      $s=4$\AA. In the contour plots  dark regions indicate low density
      regions while high densities are plotted bright. The panels
      below the contour plots show the water density $\rho$ scaled
      with water bulk density $\rho_{0}$ in a cylinder of radius
      $5$\AA, coaxial with the center-to-center line of the
      solutes.}    
\label{fig:contour_charged}
\end{center}
\end{figure}

\begin{figure}
  \begin{center}
\includegraphics[width=8cm,angle=0.,clip]{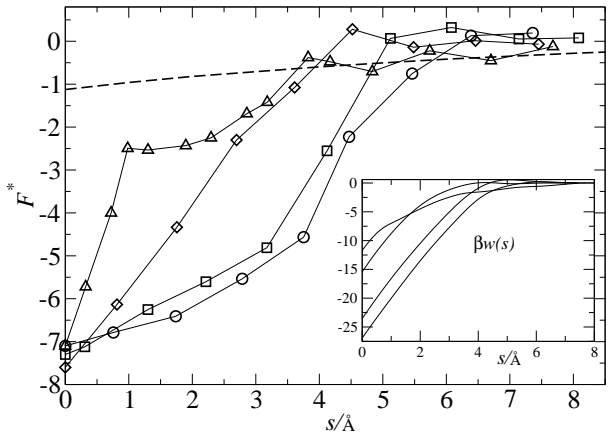}
    \caption{Mean force, $F^{*}=\beta F$\AA, as a function of
      surface-to-surface distance $s$ for neutral $\Sn$ solutes
      (circles) and oppositely charged ${\rm S}_\pm$ solutes of
      $R=11$\AA~ with different central charges $\pm qe$:
      $q=2$ (squares), $q=5$ (diamonds), $q=10$ (triangles up). The
      dashed line represents the electrostatic force between 2
      periodically repeated solutes with opposite charges $q=\pm$ 10
      in a continuous solvent with permittivity $\epsilon=80$.  The
      inset shows the resulting potentials of mean force; the contact
      values $w(s=0)$ increase with $q$.}
\label{fig:forces_opp_charged}
\end{center}
\end{figure}

\begin{figure}
  \begin{center}
\includegraphics[width=8cm,angle=0.,clip]{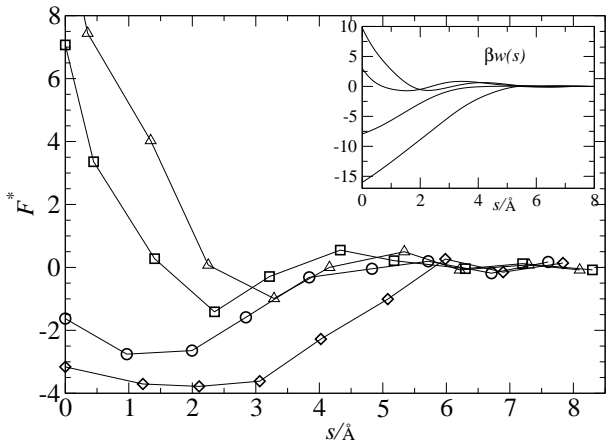}
    \caption{Mean force, $F^{*}=\beta F$\AA, as a function of
      surface-to-surface distance $s$ for equally charged ${\rm
      S}_\pm$ solutes of radius $R=11$\AA~ and $q=-4$ (diamonds),
      $q=4$ (circles), $q=-8$ (squares), and $q=8$ (triangles). The
      inset shows the according potential of mean force.}
\label{fig:forces_like_charged}
\end{center}
\end{figure}

We now turn to charged solutes. Consider first oppositely charged
solutes. The water density profiles are illustrated in
Figs.~\ref{fig:contour_charged}(a)-(d) for the case of solutes of
radius $R=11$\AA~, a surface-to-surface distance $s=4$\AA~, and
various charges $q$. The upper part of each frame shows a density
contour plot coded by variable shades of gray. The lower part shows
density profiles along the center-to-center axis $z$, averaged over a
coaxial cylindrical volume of radius $5$\AA. In Frame (a) we plot the
density distribution for the neutral case $Q=0$ for comparison with
the charged systems.  Frames (b)-(d) in Fig.~\ref{fig:contour_charged}
show water density profiles in the vicinity of two spheres carrying
opposite electric charges $\pm qe$ at their center (opposite charges
ensure overall charge neutrality without any need for counterions). As
$q$ increases from zero (frame (a)), water is seen to penetrate
between the two solutes, the central peak around $z=0$ in the density
profiles increases rapidly and its amplitude reaches roughly the bulk
density of water when $q=10$. Note that this central peak is
asymmetrically split, indicating the presence of two hydration layers
which differ somewhat depending on their association with the anionic
or cationic solute. This difference is also evident in the contact
values of the outside surfaces of the solutes, and is a consequence of
the different arrangements of the water dipoles around the solutes
induced by the local electric fields. The asymmetry of the profiles
can be rationalized by inspecting the water structure around isolated
solutes, as shown in Fig.~\ref{fig:profiles_charged}(a) and (b) and
Fig.~\ref{fig:orient} in Sec. \ref{sec:structure}. The hydration shell
is more sharply defined around the cationic than around the anionic
solute. The water dipoles tend to point radially away from the cation,
while the opposite configuration is more favorable around anions.

The resulting mean forces between solutes are plotted for $q=0,2,5$
and 10, as functions of the surface-to-surface distance $s$ in
Fig.~\ref{fig:forces_opp_charged} together with corresponding
potentials of mean force. The mean force includes the direct Coulomb
interaction between the two solutes (with proper account for the
periodic images), which is in fact an order of magnitude larger than
the total mean force. At large distances hydrophobic interactions
become negligible and the force should tend to
$-q^{2}e^{2}/(4\pi\epsilon_{0}\epsilon r^{2})$, where $r=2R+s$ and
$\epsilon$ is the dielectric permittivity of bulk water; the
corresponding curve for $q=\pm 10$ is also shown in
Fig.~\ref{fig:forces_opp_charged}.

The most striking result illustrated in
Fig.~\ref{fig:forces_opp_charged} is the near independence of the
force at contact, $s=0$, with respect to solute charge. From the
density profiles in Fig.~\ref{fig:contour_charged} the hydrophobic
attraction is expected to be reduced but this reduction is almost
exactly compensated by the Coulomb attraction between solutes in the
presence of the solvent. As $q$ increases, the initial slope of the
effective force increases. The potential of mean force (shown in the
inset to Fig.~\ref{fig:forces_opp_charged}) exhibits a contact value
which increases with $q$, indicating that the reduction of hydrophobic
attractive free energy clearly outweighs the increase in bare Coulomb attraction between the
latter. Simulations calculating the forces at and near contact for
$q=7$ and $q=15$, not shown in Fig.~\ref{fig:forces_opp_charged},
confirm this trend. Note that the potential of mean force  for
$q=10$ shows more long range attraction compared to the smaller $q$
data due to the increased electrostatic attraction. The
eye-catching kink in the force for $q=$10, at a distance $s\approx
1$\AA~ is reproducible, and is probably related to the pronounced
shell structure of water molecules around highly charged solutes,
discussed in sec~\ref{sec:structure}. While for neutral (and weakly
charged) solutes, the O and H density profiles show little structure,
they are sharply peaked at a distance $s\approx 1$\AA~ of the O atoms
from the solute surface. This would lead to a complete shared
hydration layer, and consequently to a kink in the force versus
distance curve, between 2 flat solutes separated by $s=2$\AA~. This
critical separation is shifted to shorter distances due to the
curvature of spherical solutes.
 
In view of this delicate balance between various interactions, we have
also examined the case of equally charged solutes. In this case
monovalent counterions (Na$^{+}$ or Cl$^{-}$) were included to ensure
overall charge neutrality. The situation is summarized in
Fig.~\ref{fig:forces_like_charged} for solutes of radius $R=11$\AA~
and charge $q=\pm4$ and $q=\pm8$. Charges of 4 and 8 were chosen to
allow a direct comparison with the results for the models ${\rm
T}_\pm$ and ${\rm C}_\pm$ in the next section.  The water density
profiles are symmetric with respect to $z=0$ for equally charged
solutes, but differ substantially when going from a pair of anions to
a pair of cations, as discussed in Sec.~\ref{sec:structure}. This
difference is reflected in the effective forces and potentials shown
in Fig.~\ref{fig:forces_like_charged}. The interaction between the
anionic solutes is always more attractive. For $q=\pm4$ hydrophobic
attraction overcomes the repulsion between like charges, while for
$q\pm8$ the electrostatic contribution dominates and the force is
mainly repulsive, apart from a small attractive kink at
$s=2-3$\AA. The contact value of the repulsive forces is an order of
magnitude higher than in a continuous solvent, originating obviously
from the lack of water between the solutes and, hence a reduced
dielectric screening. 

The reduction of the hydrophobic attraction between initially
uncharged solutes, upon increasing the solute charge, may be qualitatively
understood by the orienting action of the strong electric field
between charged solutes on the water molecules, which disrupts the
local hydrogen-bond structure and moves water locally away from
conditions of liquid-vapor coexistence, so that ``drying'' no longer
occurs.

\subsection{Discrete charge patterns}
\label{sec:inhom}
The water density distribution around two tetrahedral solutes $\Tn$
(i.e carrying no net charge) is plotted in Fig.~18(a)-(e) for
increasing values of the surface-to-surface distance $s$.  As in the
case of neutral solutes $\Sn$, water depletion between the solutes is
observable up to a solute distance of $s\simeq5$\AA. The bright
regions near the surfaces indicate high density water and stem from
the solvation shells in the immediate vicinity of the surface
charges. We note again that the density profiles were calculated by
averaging the water density cylindrically around the symmetry axis and
the solutes were free to rotate in the simulations. The difference in
brightness at the solute surfaces shows that certain orientational
configurations are favored over others. If the tetrahedra were
rotating freely (without any mutual interactions), the brightness
would be the same everywhere on the solute contour, as in the case of
homogeneously charged solutes. It seems that the systems chooses those
configurations where the hydrophobic parts of the solutes (i.e. areas
between the hydrophilic surface charges) face each other. We have also
performed simulations of the overall neutral tetrahedral solute
replacing the SPC/E water by a continuous solvent with permittivity
$\epsilon=80$. In the latter case and for short distances, $s\lesssim
5$\AA, configurations are favored in which opposite surface charges
associated with the two solutes face each other, thus strongly
lowering the electrostatic energy of the system. With explicit water
this is apparently no longer the case, despite the expected reduction
in dielectric screening close to the surfaces; the system tries to
deplete water from the solute surfaces between the solutes. For
distances larger than $s\simeq5$\AA~, where no visible drying
occurs, the positions of the bright regions are more smeared out on
average, indicating a higher rotational freedom of the solutes. A
simple configurational order parameter will be defined and discussed
later.  The water profiles for the tetrahedral solutes $\Tp$ show the
same behavior, in particular a visible drying for $s\lesssim5$\AA. We
have not performed simulations of a pair of the negative $\Tm$
tetrahedra but expect similar behavior.

\begin{figure}
  \begin{center}
\includegraphics[width=6cm,angle=0.,clip]{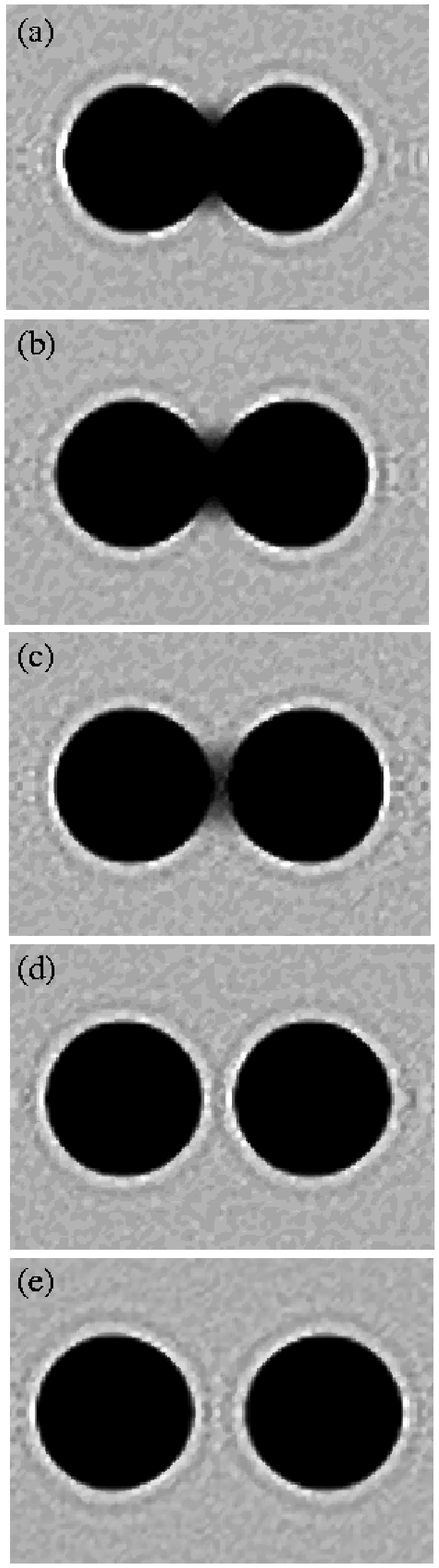}
    \caption{Contour density profiles of water around two $\Tn$
solutes for surface-to-surface distances (a) $s=0$\AA, (b) $s=2$\AA,
(c) $s=4$\AA, (d) $s=6$\AA, and (e) $s=9$\AA.}
\label{fig:contour_tetra}
\end{center}
\end{figure}

The water density profiles around a solute $\Cn$ carrying an overall
neutral cubic charge distribution are plotted in
Fig.~\ref{fig:contour_cube}(a)-(c).  No water depletion is visible at
any distance. In (a) the black region between the solutes comes from
the fact that the solute surfaces touch.  The positions of the high
density regions of water corresponding to the first solvation shells
of the surface charges are on average distributed homogeneously over
the sphere surface, pointing to a high orientational freedom of the
solutes. The density of water close to the solute surface (bright ring
in Figs.~\ref{fig:contour_cube}(a)-(c)) is on average higher compared
to the tetrahedra due to the larger surface charged density. The water
density profiles for the positive cubic solute $\Cp$ show a different
behavior, resembling the results for the tetrahedral solute, as shown
in Fig.~\ref{fig:contour_pos_cube}.  For distances $s\lesssim 4$\AA~
no bright region is found between the solutes, and hence water
depletion is observed. Similar to the tetrahedral case the solutes
stay mainly in orientational configurations in which the hydrophobic
patches face each other.

\begin{figure}
  \begin{center}
    \caption{Contour density profiles of water around two $\Cn$
solutes with cubic charge distribution for surface-to-surface
distances (a) $s=0$\AA~, (b) $s=2$\AA~, and (c) $s=4$\AA~.}
\label{fig:contour_cube}
\end{center}
\end{figure}

\begin{figure}
  \begin{center}
\includegraphics[width=7cm,angle=0.,clip]{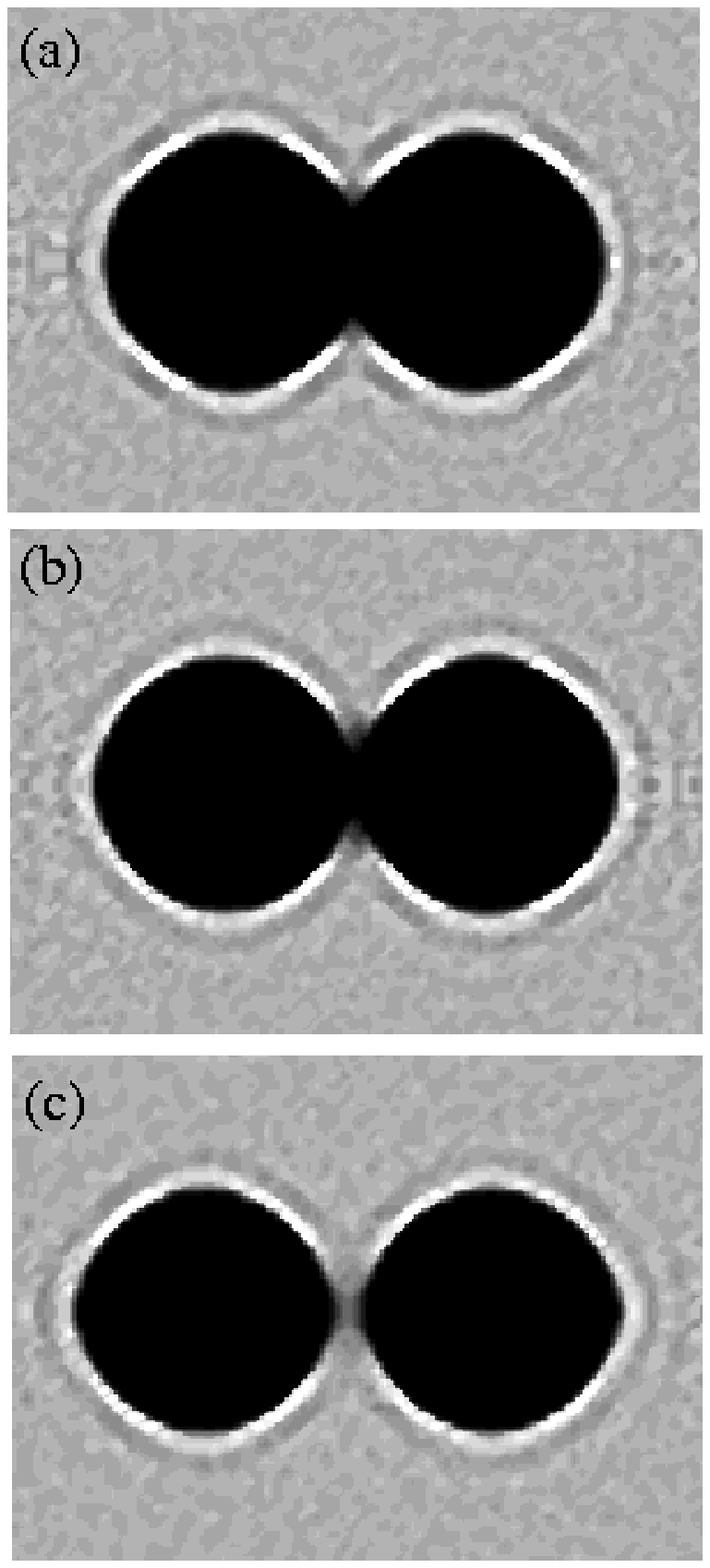}
    \caption{Contour density profiles of water around two $\Cp$
solutes for surface-to-surface distances (a) $s=0$\AA~, (b) $s=2$\AA~,
and (c) $s=4$\AA~.}
\label{fig:contour_pos_cube}
\end{center}
\end{figure}

\begin{figure}
  \begin{center}
\includegraphics[width=8cm,angle=0.,clip]{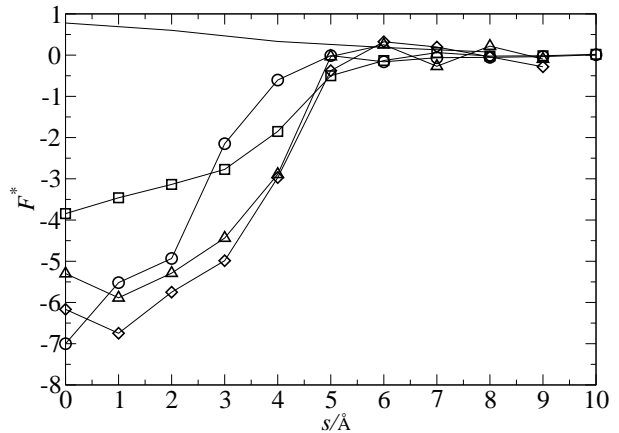}
    \caption{Mean force between  $\Tp$ (diamonds), $\Tn$ (triangles),
    $\Cp$ (squares), and $\Cn$ (circles) solutes. Also plotted
    is the force between two periodically repeated $\Cp$ solutes 
    in a continuous solvent with $\epsilon=80$ (solid line without symbols).}
\label{fig:forces_solutes}
\end{center}
\end{figure}

The effective force between the model solutes is plotted in
Fig.~\ref{fig:forces_solutes}. We show the force between pairs of
neutral and overall positive tetrahedra, $\Tn$ and $\Tp$ as well as
between pairs of neutral and overall positive cubes, $\Cn$ and $\Cp$.
Simulations with overall charged solutes were carried out with
explicit counterions.  We have not performed simulations of a pair of
overall negative tetrahedra and cubes. The force between pairs of
overall neutral and charged tetrahedra is only slightly less
attractive than between pairs of spherically symmetric neutral solutes
(compare to Fig.~\ref{fig:forces_sim}). This is surprising as one
expects a stronger influence of the high electric fields generated by
the surface charges. We learned from the investigation of
homogeneously charged solutes that an electric field can considerably
lower the hydrophobic attraction. Apparently, a more anisotropic
electric field distribution again favors hydrophobic
attraction. Compare the force between pairs of positively charged
tetrahedra and pairs of homogeneously charged solutes with the same
overall charge $q=+4$ (Fig.~\ref{fig:forces_like_charged}): the
attraction between the homogeneously charged solutes is less than half
that between $\Tp$ solutes. As already discussed above, the strong
attraction between two tetrahedra is accompanied by water depletion
between them, as in the case of homogeneous neutral solutes.  The
water density profile around an isolated tetrahedral solute ${\rm T}_\pm$,
shown in Sec. \ref{sec:structure}, already illustrated strong
depletion of water from the regions between the first solvation shells
of the discrete surface charges. A possible explanation of the strong
attraction between two tetrahedra is that this depletion is amplified
when two hydrophobic patches of the solutes come close and face each
other, and thus lowering the free energy.

The effective force between two solutes with overall neutral cubic
charge distribution shows qualitative differences compared to the
tetrahedra. Cubes with zero overall charge still attract each other,
but the interaction range is decreased.  Analysis of the
configurations shows that for close cubic solutes ($s\approx 1-3$\AA)
one positive and one negative charge belonging to different solutes
are on average very close, interacting with reduced dielectric screening
than in bulk water due to their mutual proximity. The attraction
observed is therefore mainly due to the electrostatic contribution and
not hydrophobic attraction as in the case of neutral tetrahedra. For
the $\Cp$ solute the situation is again different. All charges repel,
allowing the hydrophobic patches to face each other and water
depletion is induced, as seen in the water profiles of
Fig.~\ref{fig:contour_pos_cube}. Although equally charged, the cubic
solutes still attract each other in striking contrast to the
homogeneously charged solutes with $Q=8$ in explicit water
(Fig.~\ref{fig:forces_like_charged}) and different to the case where
water is replaced by a continuous solvent with $\epsilon=80$, also
plotted in Fig.~\ref{fig:forces_solutes}.
  
\begin{figure}
  \begin{center}
\includegraphics[width=8cm,angle=0.,clip]{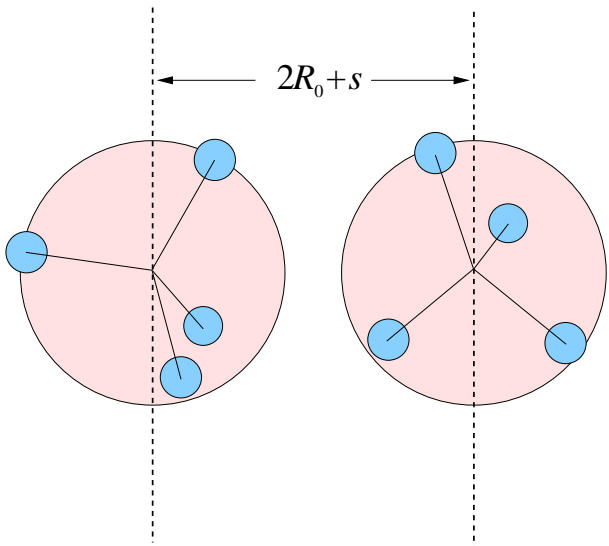}
    \caption{Sketch of two T solutes with radius $R_0$ and a 
    surface-to-surface distance $s$. The dashed lines through the
    solute centers delimit a slab of width $2R_0+s$. The numbers of
    solute charges (small spheres) of each solute $N_1$ and $N_2$
    inside the slab define the order parameter $N=N_1 \cdot N_2$,
    described in section \ref{sec:inhom}.  $N=6=3\cdot 2$ in the
    configuration shown. }
\label{fig:slab}
\end{center}
\end{figure}

\begin{figure}
  \begin{center}
\includegraphics[width=8cm,angle=0.,clip]{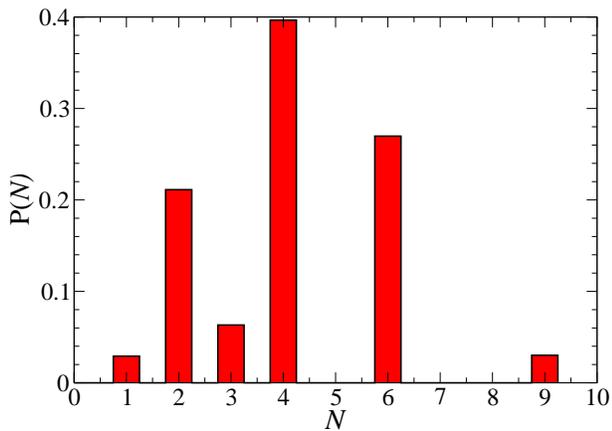}
    \caption{Configurational probabilities $P(N)$ as defined in
    Sec. \ref{sec:inhom} for two non-interacting tetrahedral solutes.}
\label{fig:prob_free}
\end{center}
\end{figure}

In the following we investigate how the explicitly resolved water
affects the average orientational configurations of two close
tetrahedral solutes, when their centers are held at fixed
positions. In a continuous solvent the probability of observing a
certain configuration is purely determined by the electrostatic
interactions between the surface charges.  Obviously, for close
distances of the solutes, structural effects of explicit water are
expected to be very significant. A simple orientational order parameter, which coarsely
probes different orientational configurations of a pair of tetrahedral
solutes can be defined as follows: we count the number of surface
charges in the slab delimited in width by the centers of the two
solutes, as sketched in Fig.~\ref{fig:slab}. Let $N_1$ and $N_2$ be
the numbers of charges in the slab belonging to the first and second
solutes. The values $N_i$, $i$=1,2 for a tetrahedron with four charged
vertices are obviously $N_i=1,2$ or 3 (it is not possible to have four
charges on one half sphere of the tetrahedral solute). The order
parameter is now defined as the product $N=N_1N_2$ and can take values
1,2,3,4,6,9, which characterizes 6 different mutual orientational
configurations. For $N=1$, for instance, one charge of each solute is
located in the slab, and the charges are both necessarily close to the
symmetry axis; on the other hand, when $N=9$, 3 charges of each solute are
within the slab and the bare triangular surfaces between the three charges are mainly
facing each other. In Fig.~\ref{fig:slab} we sketch two tetrahedra
in a configuration $N=3\cdot 2 = 6$.  The probability distribution
$P(N)$ for two freely (without any interactions) rotating tetrahedra
is plotted in Fig.~\ref{fig:prob_free}. $N=2=2\cdot 1=1\cdot 2$,
$N=4=2\cdot 2$, and $N=6=2\cdot 3=3\cdot 2$ are the most likely
configurations, with probabilities $P(2)\simeq0.21$, $P(4)=\simeq
0.4$, and $P(6)=0.28$.

\begin{figure}
  \begin{center}
\includegraphics[width=8cm,angle=0.,clip]{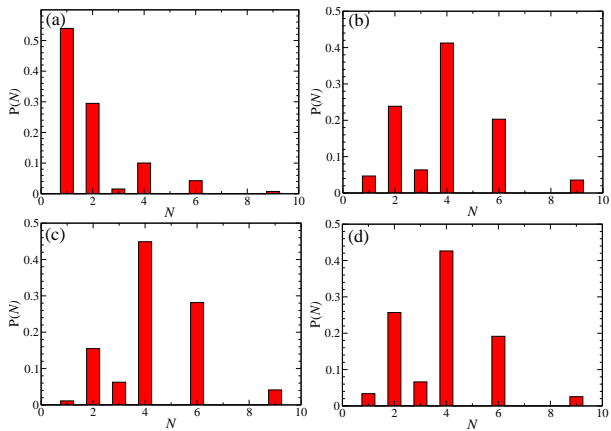}
    \caption{Configurational probabilities $P(N)$ as defined in
    Sec. \ref{sec:inhom} for two tetrahedral solutes in a continuous solvent
    with $\epsilon=80$. (a) and (b) are for $\Tn$ solutes at
    a surface-to-surface distance  $s=3$\AA~ and $s=9$\AA. (c)
    and (d) are for $\Tp$ solutes at a distance
    $s=3$\AA~ and $s=9$\AA, resp.}
\label{fig:prob_eps80}
\end{center}
\end{figure}

In Fig.~\ref{fig:prob_eps80}(a)-(d) we plot the probability
distribution $P(N)$ for interacting tetrahedra in a continuous solvent
with permittivity $\epsilon=80$.  In Fig.~\ref{fig:prob_eps80}(a) and
(b) we show the result for a pair of overall neutral $\Tn$ solutes at
distances $s=3$\AA~ and $s=9$\AA~.  For the close distance the free
rotator distribution is dramatically changed and the $N=1$ and $N=2$
configurations are strongly favored. This is due to negative and
positive charges from different solutes attracting each other at close
distance. For the larger distance the electrostatic interactions are
weaker and $P(N)$ strongly resembles the free rotator distribution
again.  In Fig.~\ref{fig:prob_eps80}(c) and (d) we show the same
distribution function, now for a pair of overall positive solutes at
distances $s=3$\AA~ and $s=9$\AA~, resp. Here, at close distance the
$N=1$ and $N=2$ configurations are suppressed due to the proximity of
like charges, and the $N>3$ configurations are enhanced, since they
allow the charges of one solute to be at larger mean distance from the
like charges of the second solute. For large distances, (d), we again
recover the free rotator distribution.

\begin{figure}
  \begin{center}
\includegraphics[width=8cm,angle=0.,clip]{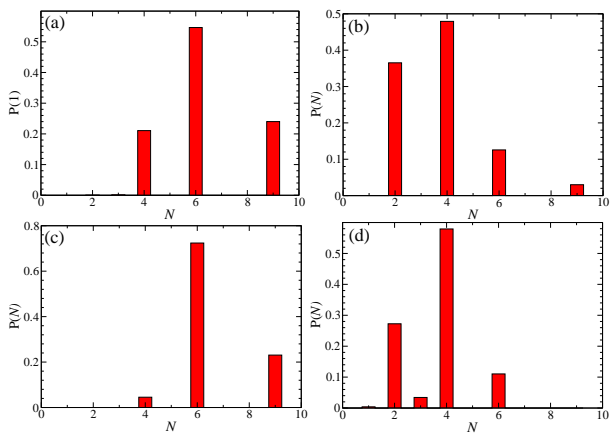}
    \caption{Same as in Fig.~\ref{fig:prob_eps80}, but now with
    explicit SPC/E water instead of a continuous solvent.}
\label{fig:prob_exp}
\end{center}
\end{figure}

In Fig.~\ref{fig:prob_exp}(a)-(d) the continuous solvent is now
replaced by explicit water molecules. For close solutes ($s=3$\AA) the
difference with the continuous solvent is large: the probabilities of
the $N=1$ and $N=2$ configurations are reduced both for the neutral
(a) and overall charged tetrahedra (c). The $N=6$ and $N=9$
configurations are greatly enhanced, indicating that on average the
are hydrophobic surfaces face each other, as expected from the water profiles in Fig.~\ref{fig:contour_tetra}(b),(c). Remarkably,
even for case (a) the explicit solvent system strongly favors water
depletion rather than the proximity of two unlike charges, which would
lower the electrostatic energy significantly. Increasing the distance to
$s=9$\AA, the probabilities of the $N=6$ and $N=9$ configurations are
lowered and the overall distributions, both for neutral and positive
tetrahedra are more similar to the free rotator distribution.
\vspace{-1cm}
\section{Conclusion}
\label{sec:conclusion}
We have used a simple model of neutral and charged, nanometer-sized
spherical solutes, embedded in explicit aqueous solvent, to
investigate the influence of charge patterns on the solvation of a
single solute, and on the effective, solvent-induced interaction
between two solutes. The charge patterns considered in this paper
include uniform charge distributions (equivalent to a single charge at
the center of the spherical solute), as well as tetrahedral or cubic
charge distributions, involving 4 or 8 discrete positive or negative
charges situated at the solute surface, adding up to an overall
positive, zero or negative charge $Q$ (the $\Tp,\Tm,\Tn$ and
$\Cp,\Cm,\Cn$ models).
 
Extensive constant pressure and constant temperature ($NPT$) Molecular
Dynamics simulations were carried out under ``normal'' solvent
conditions, i.e. close to liquid-vapor coexistence of water at room
temperature. These simulations provide water density profiles around a
single solute or a pair of solutes, which can be resolved into
solute-oxygen and solute-hydrogen pair distribution functions, the
distance resolved orientational order parameter $P(r)$, the solvation
free energy as a function of solute radius and charge, as well as the
effective force and pair potential between two solutes, averaged over
solvent configurations. The main results of this investigation may be
summarized as follows:

1. The density profiles of water around a single, neutral solute
   ($\Sn$ model), and their variation with solute radius $R$
   (cf. Fig.~3) exhibit the characteristic ``destructuring'' for radii
   $R\gtrsim 5$\AA~ already reported by earlier studies.
   \cite{stilinger,huang:jpc,lum:jpc} The water molecules exhibit no
   significant orientational ordering around neutral nano-sized solutes.

2. The hydrogen and oxygen density profiles change dramatically when
   the solute is uniformly charged (${\rm S}_\pm$ models).  These
   profiles are sensitive to the anionic or cationic nature of the
   solute (for a given absolute charge $|Q|$), in addition the
   hydrogen profiles exhibit a splitting of the main peak in the case
   of anionic ($\Sm$) solutes (cf. Fig.~5). The orientational order
   parameter $P(r)$ exhibits a significant structure, and a relatively
   slow decay with $r$, indicative of strong orientational ordering
   around the solutes $\Sp$ or $\Sm$, which is somewhat more
   pronounced around an anionic solute. The hydration asymmetry
   results in preferential solvation of anionic solutes for a given
   radius and absolute charge $|Q|$, in agreement with earlier
   findings. \cite{hummer:jpc:1996,lynden-bell,garde:2004}

3. Moving from uniformly charged solutes to discrete (tetrahedral or
   cubic) charge patterns, the hydration of nano-sized solutes is found to
   exhibit a strong angular modulation associated with the hydrophilic
   ``patches'' around the discrete surface charges, and hydrophobic
   ``patches'' in between (cf. Fig. 6). The conflicting hydration
   patterns lead to a surprising depletion of water around $\Tp$ or
   $\Tm$ solutes, compared to a neutral solute $\Sn$. The solvation
   free energy is found to be about 20 $\%$ lower for solutes with
   discrete charge patterns compared to that of uniformly charged
   solutes with the same overall charge (cf. Fig. 9).

4. The present simulations confirm the strong hydrophobic attraction
   between two neutral spherical nano-sized solutes linked to solvent
   ``drying'', which was already reported earlier for similar
   models. The MD results for solute radii $R\gtrsim 5$\AA~ are nearly
   quantitatively reproduced by a simple calculation based on purely
   macroscopic considerations, and the force at solute-solute contact
   is found to scale roughly linearly with $R$.

5. The effective attraction between neutral solutes is strongly
   reduced, or turns into a repulsion, when the nano-sized solutes carry
   equal, uniform charge distributions. The total force has a
   repulsive electrostatic component, while examination of the water
   density profiles shows that the ``drying'' is mostly
   suppressed. The effective force is systematically less attractive
   (or more repulsive) between pairs of cationic solutes compared to
   anionic pairs (cf. Fig. 17). Turning to a pair of oppositely (but
   uniformly) charged solutes, the MD simulations show that the range
   of the effective attraction decreases when the absolute charge
   $|Q|$ increases, again in agreement with simple macroscopic
   considerations, but that the effective force at contact seems to be
   independent of $Q$, and equal to the hydrophobic force between
   neutral solutes; we have no explanation for this surprising
   observation.

6. The situation for discrete solute charge patterns is, not
   surprisingly, more complex, due to the competition between the
   resulting hydrophilic and hydrophobic ``patches'' on the solute
   surface. On average, some ``drying'' of water is observed, and the
   resulting mean force between solutes carrying tetrahedral or cubic
   patterns is once more attractive, despite the electrostatic
   repulsion (``like-charge attraction''). This effect is obviously
   incompatible with crude ``implicit solvent'' models.

7.  The complete break-down of ``implicit solvent'' models, whereby
the latter is replaced by a dielectric continuum, is further
illustrated by the highly coarse-grained representation of the
configurational probability density of two solutes carrying discrete
charge distributions, introduced in Sec.~V. The relative orientations
of the surface charge patterns on the two solutes are completely
different for explicit and implicit solvent models, particularly at
short surface-to-surface distances $s$ (cf. Figs. 24 and 25).

The key message of the present work is that explicit solvent models
are unavoidable for a proper description of the effective interactions
between nano-sized solutes like proteins, and that the latter are extremely
sensitive to the precise location of any electric charges carried by
the solutes. Contrarily to effective interactions on larger colloidal
scales, a generic coarse-graining strategy appears to be useless when
solutes in the nanometer range are considered, and fully molecular
models are required for realistic simulations.

\section{Acknowledgments}
JD acknowledges the financial support of EPSRC within the Portfolio
Grant RG37352. We thank A. Archer and V. Ballenegger for useful
discussions, and A.A. Louis for using the computer cluster Ice.

\section{Appendix A: Simulation details}
The simulations were performed with the DLPOLY2 \cite{dlpoly}
package. The Berendsen barostat and thermostat \cite{berendsen:jcp}
were used to maintain the SPC/E water at a pressure of 1 bar and a
temperature $T=300$K.  For the simulations of the solutes with
inhomogeneous charge distributions (T and C models) we used the rigid
body algorithm with quaternions to properly account for the rotation
of the anisotropic solutes. To this end we switched to an integration
routine using the Nos{\'e}-Hoover barostat and thermostat, which
turned out to be more stable in conjunction with quaternions. We
carefully checked that both barostats give the same results by
performing tests with bulk water, treated both with bond contraints
and with the rigid body algorithm. Test runs using the Nos{\'e}-Hoover
barostat and thermostat for the S models also showed no difference.

The simulation cell is a periodically repeated cube with a maximum
boxlength of about $L=48$\AA, containing up to $N_{\rm w}$=3000 water
molecules, depending on the solute size. For simulations of one
isolated solute we required that the surface-to-surface distance to
the nearest image solute was 20\AA, yielding a box size of
$L=2R+20$\AA. For the calculation of the interaction force between two
solutes, the latter are placed at fixed positions ${{\vec R}_1}$ and
${{\vec R}_2}$ on the body diagonal of the simulation cell. The center
to center distance is then ${R_{12}}={|{\vec R}_1}-{{\vec
R}_2}|$.  The corresponding box dimensions are chosen such that the
surface-to-surface distance $s={R_{12}}-2R_{0}$ to the nearest image
solute is 20\AA. The box length can be calculated as $L=(4R+s+20{\rm
\AA})/\sqrt{3}$.  Due to the constant pressure constraint the box length
fluctuates slightly in the simulations. The long range electrostatic
interactions were evaluated with smooth particle-mesh Ewald (SPME)
summations \cite{essmann:jcp} using 16 $\vec k$ vectors in each
direction and a convergence parameter of 3.2$/r_{\rm cut}$. A
cutoff distance $r_{\rm cut}=9$\AA~ was used for LJ-interactions and
the real space SPME contributions. For the nanosized solutes
a larger cut-off radius is obviously required. We optimize the computational
speed by introducing a second cutoff for the solute-water and
solute-ion interactions, chosen to be $R_0+4$\AA, sufficient large for
the shifted, short ranged repulsive interaction~(\ref{eq:solutepot}).

\section{Appendix B: Finite corrections for solvation free energies from simulation}
Accurate solavtion free energies for charged
solutes can be obtained by Ewald summations in periodic cells\cite{hansen} when the
self-interacion energy of the solute charges with its periodic images
and the background charge is properly included.\cite{hummer:jpc:1996}
This correction is slightly modified when the solute size $R$ is
comparable to the box size $L$.\cite{hummer:jcp:1997} The final
expression for the electrostatic contribution to the solvation free
energy for our solute models, including the finite size corrections,
is:
\begin{eqnarray}
\Delta\mu_\pm &=& \Delta\mu_\pm^{\rm sim} \cr & + &
 \frac{e^2}{8\pi\epsilon_0}\frac{\epsilon-1}{\epsilon} \left(\frac{\xi_{\rm
 EW}}{L}\sum_{\alpha=1}^{N_{\rm c}}q_{\alpha}^{2}+\frac{2\pi
 R^2}{3L^3}\sum_{\alpha=1}^{N_{\rm c}}q_{\alpha}^2\right) \cr
 &+&\frac{e^2}{8\pi\epsilon_0}\frac{\epsilon-1}{\epsilon}\sum_{\alpha=1}^{N_{\rm
 c}}\sum_{\beta\neq\alpha,1}^{N_{\rm c}}q_{\alpha} q_{\beta} \left(\phi_{\rm
 EW}(\vec r_{\alpha\beta})-\frac{1}{|\vec r_{\alpha\beta}|}\right),\nonumber \\
\label{eq:correct}
\end{eqnarray}
where $\epsilon$ is the macroscopic permittivity of water, and for a
periodic array of cubic simulation cells,
$\xi_{EW}\approx-2.837297$. In the case of a uniformly charged
solute corresponding to a single charged site $qe$ at the center, the
result (\ref{eq:correct}) reduces to\cite{hummer:jcp:1997}
\begin{eqnarray}
\Delta\mu_{\pm}=\Delta\mu_\pm^{\rm sim}+\frac{q^2 e^2}{8\pi\epsilon_0}\frac{\epsilon-1}{\epsilon}\left(\frac{\xi_{\rm EW}}{L}+\frac{2\pi R^2}{3L^3}\right).
\label{eq:correct2}
\end{eqnarray}
With the system sizes used in the present simulation the finite size
corrections are very large and represent typically twice the value of
$\Delta\mu_{\rm sim}$ and stem mainly from the $\xi_{\rm EW}$-term in
Eqs.~(19) and (20).


\begin{thebibliography}{29}
\expandafter\ifx\csname natexlab\endcsname\relax\def\natexlab#1{#1}\fi
\expandafter\ifx\csname bibnamefont\endcsname\relax
  \def\bibnamefont#1{#1}\fi
\expandafter\ifx\csname bibfnamefont\endcsname\relax
  \def\bibfnamefont#1{#1}\fi
\expandafter\ifx\csname citenamefont\endcsname\relax
  \def\citenamefont#1{#1}\fi
\expandafter\ifx\csname url\endcsname\relax
  \def\url#1{\texttt{#1}}\fi
\expandafter\ifx\csname urlprefix\endcsname\relax\def\urlprefix{URL }\fi
\providecommand{\bibinfo}[2]{#2}
\providecommand{\eprint}[2][]{\url{#2}}

\bibitem[{\citenamefont{Rubinstein and Colby}(2003)}]{rubinstein}
\bibinfo{author}{\bibfnamefont{M.}~\bibnamefont{Rubinstein}} \bibnamefont{and}
  \bibinfo{author}{\bibfnamefont{R.~H.} \bibnamefont{Colby}},
  \emph{\bibinfo{title}{Polymer Physics}} (\bibinfo{publisher}{Oxford
  University Press}, \bibinfo{year}{2003}).

\bibitem[{\citenamefont{Chandler}(2004)}]{chandler_review}
\bibinfo{author}{\bibfnamefont{D.}~\bibnamefont{Chandler}}
  (\bibinfo{year}{2004}), \bibinfo{note}{to appear in Nature}.

\bibitem[{\citenamefont{Likos}(2001)}]{likos:physrep}
\bibinfo{author}{\bibfnamefont{C.~N.} \bibnamefont{Likos}},
  \bibinfo{journal}{Phys. Rep.} \textbf{\bibinfo{volume}{348}},
  \bibinfo{pages}{267} (\bibinfo{year}{2001}).

\bibitem[{\citenamefont{Stilinger}(1973)}]{stilinger}
\bibinfo{author}{\bibfnamefont{F.~H.} \bibnamefont{Stilinger}},
  \bibinfo{journal}{J. Solution Chem.} \textbf{\bibinfo{volume}{2}},
  \bibinfo{pages}{141} (\bibinfo{year}{1973}).

\bibitem[{\citenamefont{Hummer and Garde}(1998)}]{hummer:prl}
\bibinfo{author}{\bibfnamefont{G.}~\bibnamefont{Hummer}} \bibnamefont{and}
  \bibinfo{author}{\bibfnamefont{S.}~\bibnamefont{Garde}},
  \bibinfo{journal}{Phys. Rev. Lett.} \textbf{\bibinfo{volume}{80}},
  \bibinfo{pages}{4193} (\bibinfo{year}{1998}).

\bibitem[{\citenamefont{Lum et~al.}(1999)\citenamefont{Lum, Chandler, and
  Weeks}}]{lum:jpc}
\bibinfo{author}{\bibfnamefont{K.}~\bibnamefont{Lum}},
  \bibinfo{author}{\bibfnamefont{D.}~\bibnamefont{Chandler}}, \bibnamefont{and}
  \bibinfo{author}{\bibfnamefont{J.~D.} \bibnamefont{Weeks}},
  \bibinfo{journal}{J. Phys. Chem. B} \textbf{\bibinfo{volume}{103}},
  \bibinfo{pages}{4570} (\bibinfo{year}{1999}).

\bibitem[{\citenamefont{Huang et~al.}(2001)\citenamefont{Huang, Geissler, and
  Chandler}}]{huang:jpc}
\bibinfo{author}{\bibfnamefont{D.~M.} \bibnamefont{Huang}},
  \bibinfo{author}{\bibfnamefont{P.~L.} \bibnamefont{Geissler}},
  \bibnamefont{and} \bibinfo{author}{\bibfnamefont{D.}~\bibnamefont{Chandler}},
  \bibinfo{journal}{J. Phys. Chem. B} \textbf{\bibinfo{volume}{105}},
  \bibinfo{pages}{6704} (\bibinfo{year}{2001}).

\bibitem[{\citenamefont{Wallquist and Berne}(1995)}]{wallquist:jpc}
\bibinfo{author}{\bibfnamefont{A.}~\bibnamefont{Wallquist}} \bibnamefont{and}
  \bibinfo{author}{\bibfnamefont{B.~J.} \bibnamefont{Berne}},
  \bibinfo{journal}{J. Phys. Chem.} \textbf{\bibinfo{volume}{99}},
  \bibinfo{pages}{2893} (\bibinfo{year}{1995}).

\bibitem[{\citenamefont{Shinto et~al.}(1999)\citenamefont{Shinto, Miyahara, and
  Higashitani}}]{shinto}
\bibinfo{author}{\bibfnamefont{H.}~\bibnamefont{Shinto}},
  \bibinfo{author}{\bibfnamefont{M.}~\bibnamefont{Miyahara}}, \bibnamefont{and}
  \bibinfo{author}{\bibfnamefont{K.}~\bibnamefont{Higashitani}},
  \bibinfo{journal}{J. Coll. Interface Sci.} \textbf{\bibinfo{volume}{209}},
  \bibinfo{pages}{79} (\bibinfo{year}{1999}).

\bibitem[{\citenamefont{Kinoshita et~al.}(1996)\citenamefont{Kinoshita, Iba,
  Kuwamoto, and Harada}}]{kinoshita}
\bibinfo{author}{\bibfnamefont{M.}~\bibnamefont{Kinoshita}},
  \bibinfo{author}{\bibfnamefont{S.}~\bibnamefont{Iba}},
  \bibinfo{author}{\bibfnamefont{K.}~\bibnamefont{Kuwamoto}}, \bibnamefont{and}
  \bibinfo{author}{\bibfnamefont{M.}~\bibnamefont{Harada}},
  \bibinfo{journal}{J. Chem. Phys.} \textbf{\bibinfo{volume}{105}},
  \bibinfo{pages}{7177} (\bibinfo{year}{1996}).

\bibitem[{\citenamefont{Qin and Fichthorn}(2003)}]{qin}
\bibinfo{author}{\bibfnamefont{Y.}~\bibnamefont{Qin}} \bibnamefont{and}
  \bibinfo{author}{\bibfnamefont{K.~A.} \bibnamefont{Fichthorn}},
  \bibinfo{journal}{J. Chem. Phys.} \textbf{\bibinfo{volume}{119}},
  \bibinfo{pages}{9745} (\bibinfo{year}{2003}).

\bibitem[{\citenamefont{Allahyarov et~al.}(2002)\citenamefont{Allahyarov,
  L{\"o}wen, Louis, and Hansen}}]{elshad:epl:2002}
\bibinfo{author}{\bibfnamefont{E.}~\bibnamefont{Allahyarov}},
  \bibinfo{author}{\bibfnamefont{H.}~\bibnamefont{L{\"o}wen}},
  \bibinfo{author}{\bibfnamefont{A.}~\bibnamefont{Louis}}, \bibnamefont{and}
  \bibinfo{author}{\bibfnamefont{J.}~\bibnamefont{Hansen}},
  \bibinfo{journal}{Europhys. Lett.} \textbf{\bibinfo{volume}{57}},
  \bibinfo{pages}{731} (\bibinfo{year}{2002}).

\bibitem[{\citenamefont{Berendsen et~al.}(1987)\citenamefont{Berendsen,
  Grigera, and Straatsma}}]{berendsen:jpc}
\bibinfo{author}{\bibfnamefont{H.~J.~C.} \bibnamefont{Berendsen}},
  \bibinfo{author}{\bibfnamefont{J.~R.} \bibnamefont{Grigera}},
  \bibnamefont{and} \bibinfo{author}{\bibfnamefont{T.~P.}
  \bibnamefont{Straatsma}}, \bibinfo{journal}{J. Phys. Chem.}
  \textbf{\bibinfo{volume}{91}}, \bibinfo{pages}{6269} (\bibinfo{year}{1987}).

\bibitem[{\citenamefont{Dzubiella and Hansen}(2003)}]{dzubiella:jcp:2003}
\bibinfo{author}{\bibfnamefont{J.}~\bibnamefont{Dzubiella}} \bibnamefont{and}
  \bibinfo{author}{\bibfnamefont{J.-P.} \bibnamefont{Hansen}},
  \bibinfo{journal}{J. Chem. Phys.} \textbf{\bibinfo{volume}{119}},
  \bibinfo{pages}{12049} (\bibinfo{year}{2003}).

\bibitem[{\citenamefont{Spohr}(1999)}]{spohr:1999}
\bibinfo{author}{\bibfnamefont{E.}~\bibnamefont{Spohr}},
  \bibinfo{journal}{Electrochim. Acta} \textbf{\bibinfo{volume}{44}},
  \bibinfo{pages}{1697} (\bibinfo{year}{1999}).

\bibitem[{\citenamefont{Essmann et~al.}(1995)\citenamefont{Essmann, Perera,
  Berkowitz, Darden, Lee, and Pedersen}}]{essmann:jcp}
\bibinfo{author}{\bibfnamefont{U.}~\bibnamefont{Essmann}},
  \bibinfo{author}{\bibfnamefont{L.}~\bibnamefont{Perera}},
  \bibinfo{author}{\bibfnamefont{M.~L.} \bibnamefont{Berkowitz}},
  \bibinfo{author}{\bibfnamefont{T.}~\bibnamefont{Darden}},
  \bibinfo{author}{\bibfnamefont{H.}~\bibnamefont{Lee}}, \bibnamefont{and}
  \bibinfo{author}{\bibfnamefont{L.~G.} \bibnamefont{Pedersen}},
  \bibinfo{journal}{J. Chem. Phys.} \textbf{\bibinfo{volume}{103}},
  \bibinfo{pages}{8577} (\bibinfo{year}{1995}).

\bibitem[{\citenamefont{Smith and Forester}(1999)}]{dlpoly}
\bibinfo{author}{\bibfnamefont{W.}~\bibnamefont{Smith}} \bibnamefont{and}
  \bibinfo{author}{\bibfnamefont{T.~R.} \bibnamefont{Forester}}
  (\bibinfo{year}{1999}), \bibinfo{note}{the DLPOLY\_2 User Manual}.

\bibitem[{\citenamefont{Frenkel and Smit}(1996)}]{frenkelsmit}
\bibinfo{author}{\bibfnamefont{D.}~\bibnamefont{Frenkel}} \bibnamefont{and}
  \bibinfo{author}{\bibfnamefont{B.}~\bibnamefont{Smit}},
  \emph{\bibinfo{title}{Understanding Molecular Simulation: From Algorithms to
  Applications}} (\bibinfo{publisher}{Academic Press}, \bibinfo{year}{1996}).

\bibitem[{\citenamefont{Born}(1920)}]{born}
\bibinfo{author}{\bibfnamefont{M.}~\bibnamefont{Born}}, \bibinfo{journal}{Z.
  Phys.} \textbf{\bibinfo{volume}{1}}, \bibinfo{pages}{45}
  (\bibinfo{year}{1920}).

\bibitem[{\citenamefont{Reiss et~al.}(1959)\citenamefont{Reiss, Frisch, and
  Lebowitz}}]{reiss:jcp:1959}
\bibinfo{author}{\bibfnamefont{H.}~\bibnamefont{Reiss}},
  \bibinfo{author}{\bibfnamefont{H.~L.} \bibnamefont{Frisch}},
  \bibnamefont{and} \bibinfo{author}{\bibfnamefont{J.~L.}
  \bibnamefont{Lebowitz}}, \bibinfo{journal}{J. Chem. Phys.}
  \textbf{\bibinfo{volume}{31}}, \bibinfo{pages}{369} (\bibinfo{year}{1959}).

\bibitem[{\citenamefont{Latimer et~al.}(1939)\citenamefont{Latimer, Pitzer, and
  Slansky}}]{latimer:jcp:1939}
\bibinfo{author}{\bibfnamefont{W.~M.} \bibnamefont{Latimer}},
  \bibinfo{author}{\bibfnamefont{K.~S.} \bibnamefont{Pitzer}},
  \bibnamefont{and} \bibinfo{author}{\bibfnamefont{C.~M.}
  \bibnamefont{Slansky}}, \bibinfo{journal}{J. Chem. Phys.}
  \textbf{\bibinfo{volume}{7}}, \bibinfo{pages}{108} (\bibinfo{year}{1939}).

\bibitem[{\citenamefont{Hummer et~al.}(1996)\citenamefont{Hummer, Pratt, and
  Garcia}}]{hummer:jpc:1996}
\bibinfo{author}{\bibfnamefont{G.}~\bibnamefont{Hummer}},
  \bibinfo{author}{\bibfnamefont{L.}~\bibnamefont{Pratt}}, \bibnamefont{and}
  \bibinfo{author}{\bibfnamefont{A.~E.} \bibnamefont{Garcia}},
  \bibinfo{journal}{J. Phys. Chem.} \textbf{\bibinfo{volume}{100}},
  \bibinfo{pages}{1206} (\bibinfo{year}{1996}).

\bibitem[{\citenamefont{Hummer et~al.}(1997)\citenamefont{Hummer, Pratt, and
  Garcia}}]{hummer:jcp:1997}
\bibinfo{author}{\bibfnamefont{G.}~\bibnamefont{Hummer}},
  \bibinfo{author}{\bibfnamefont{L.}~\bibnamefont{Pratt}}, \bibnamefont{and}
  \bibinfo{author}{\bibfnamefont{A.~E.} \bibnamefont{Garcia}},
  \bibinfo{journal}{J. Chem. Phys.} \textbf{\bibinfo{volume}{107}},
  \bibinfo{pages}{9275} (\bibinfo{year}{1997}).

\bibitem[{\citenamefont{Lynden-Bell and Rasaiah}(1997)}]{lynden-bell}
\bibinfo{author}{\bibfnamefont{R.~M.} \bibnamefont{Lynden-Bell}}
  \bibnamefont{and} \bibinfo{author}{\bibfnamefont{J.~C.}
  \bibnamefont{Rasaiah}}, \bibinfo{journal}{J. Chem. Phys.}
  \textbf{\bibinfo{volume}{107}}, \bibinfo{pages}{1981} (\bibinfo{year}{1997}).

\bibitem[{\citenamefont{Rajamani et~al.}(2004)\citenamefont{Rajamani, Ghosh,
  and Garde}}]{garde:2004}
\bibinfo{author}{\bibfnamefont{S.}~\bibnamefont{Rajamani}},
  \bibinfo{author}{\bibfnamefont{T.}~\bibnamefont{Ghosh}}, \bibnamefont{and}
  \bibinfo{author}{\bibfnamefont{S.}~\bibnamefont{Garde}}, \bibinfo{journal}{J.
  Chem. Phys.} \textbf{\bibinfo{volume}{120}}, \bibinfo{pages}{4457}
  (\bibinfo{year}{2004}).

\bibitem[{\citenamefont{Bolhuis and Chandler}(2000)}]{bolhuis}
\bibinfo{author}{\bibfnamefont{P.~G.} \bibnamefont{Bolhuis}} \bibnamefont{and}
  \bibinfo{author}{\bibfnamefont{D.}~\bibnamefont{Chandler}},
  \bibinfo{journal}{J. Chem. Phys.} \textbf{\bibinfo{volume}{113}},
  \bibinfo{pages}{8154} (\bibinfo{year}{2000}).

\bibitem[{\citenamefont{Louis et~al.}(2002)\citenamefont{Louis, Bolhuis, and
  Hansen}}]{louis:jcp}
\bibinfo{author}{\bibfnamefont{A.~A.} \bibnamefont{Louis}},
  \bibinfo{author}{\bibfnamefont{P.~G.} \bibnamefont{Bolhuis}},
  \bibnamefont{and} \bibinfo{author}{\bibfnamefont{J.~P.}
  \bibnamefont{Hansen}}, \bibinfo{journal}{J. Chem. Phys.}
  \textbf{\bibinfo{volume}{117}}, \bibinfo{pages}{1893} (\bibinfo{year}{2002}).

\bibitem[{\citenamefont{Berendsen et~al.}(1984)\citenamefont{Berendsen, Postma,
  van Gunsteren, DiNola, and Haak}}]{berendsen:jcp}
\bibinfo{author}{\bibfnamefont{H.~J.~C.} \bibnamefont{Berendsen}},
  \bibinfo{author}{\bibfnamefont{J.~P.~M.} \bibnamefont{Postma}},
  \bibinfo{author}{\bibfnamefont{W.~F.} \bibnamefont{van Gunsteren}},
  \bibinfo{author}{\bibfnamefont{A.}~\bibnamefont{DiNola}}, \bibnamefont{and}
  \bibinfo{author}{\bibfnamefont{J.~R.} \bibnamefont{Haak}},
  \bibinfo{journal}{J. Chem. Phys.} \textbf{\bibinfo{volume}{81}},
  \bibinfo{pages}{3684} (\bibinfo{year}{1984}).

\bibitem[{\citenamefont{Hansen}(1986)}]{hansen}
\bibinfo{author}{\bibfnamefont{J.-P.} \bibnamefont{Hansen}},
  \emph{\bibinfo{title}{Molecular Dynamics Simulation of Statistical Mechanical
  Systems}} (\bibinfo{publisher}{North Holland}, \bibinfo{address}{Amsterdam},
  \bibinfo{year}{1986}), \bibinfo{note}{edited by G. Ciccotti and W.G. Hoover}.

\end{thebibliography}
\end{document}